\documentclass[aps,prl,twocolumn,longbibliography,superscriptaddress,floatfix]{revtex4-1}

\usepackage[T1]{fontenc} 

\usepackage[unicode,colorlinks]{hyperref}
\usepackage{amsmath,amssymb,amsfonts}
\usepackage{bbm}         
\usepackage{graphicx}
\usepackage{physics}
\usepackage[dvipsnames]{xcolor}
\usepackage[english]{babel}

\newcommand{\Id}{\mathbbm{1}}
\renewcommand{\Im}{\mathop{\text{Im}}}

\newcommand{\Z}{\mathbb{Z}}

\renewcommand{\>}{\rangle}
\newcommand{\<}{\langle}

\DeclareMathOperator{\sgn}{sgn}
\DeclareMathOperator{\diag}{diag}
\newcommand{\eff}{\text{eff}}

\let\appendixsection\subsection
\renewcommand{\subsection}[1]{\textit{#1.---}}
\renewcommand{\section}[1]{\textit{#1.---}}

\begin{document}

\title{Bulk-boundary correspondence for non-Hermitian Hamiltonians
\\via Green functions
}

\author{Heinrich-Gregor Zirnstein}
\affiliation{Institut f\"{u}r Theoretische Physik, Universit\"{a}t Leipzig, Br\"{u}derstrasse 16, 04103 Leipzig, Germany}
\author{Gil Refael}
\affiliation{Institute of Quantum Information and Matter and Department of Physics,
California Institute of Technology, Pasadena, CA 91125, USA}
\author{Bernd Rosenow}
\affiliation{Institut f\"{u}r Theoretische Physik, Universit\"{a}t Leipzig, Br\"{u}derstrasse 16, 04103 Leipzig, Germany}
\affiliation{Department of Condensed Matter Physics, Weizmann Institute of Science, Rehovot 76100, Israel}
\date{July 16, 2020}

\begin{abstract}
\normalsize
Genuinely non-Hermitian topological phases can be realized in open systems with sufficiently strong gain and loss; in such phases, the Hamiltonian cannot be deformed into a gapped Hermitian Hamiltonian without energy bands touching each other.
Comparing Green functions for periodic and open boundary conditions we find that, in general, there is no correspondence between topological invariants computed for periodic boundary conditions, and boundary eigenstates observed for open boundary conditions.
Instead, we find that the non-Hermitian winding number in one dimension signals a topological phase transition in the bulk: It implies spatial growth of the bulk Green function. 
\end{abstract}

\maketitle

%
Topology has made a profound impact on the description and design of wave-like systems such as quantum mechanical electrons~\cite{Hasan:2010a,Qi:2011,Chiu:2016a,Bernevig:2013,Bansil:2016} or light interacting with matter~\cite{Raghu:2008,Lu:2014,Khanikaev:2017,Rechtsman:2013,Plotnik:2014,Noh:2018}.
The key idea is to group physical systems, each described by a gapped (insulating) Hamiltonian, into the same topological class if their Hamiltonians can be continuously deformed into each other without closing the energy gap.
For Hermitian Hamiltonians, the \emph{bulk-boundary correspondence} states that topological invariants for periodic boundary conditions predict the presence of boundary states for open boundary conditions~\cite{Halperin:1982,Hatsugai:1993,Hasan:2010a,Essin:2011,Graf:2013,Avila:2013}.

Recently, non-Hermitian Hamiltonians~\cite{Hatano:1996,Bender:1998,Rotter:2009,Cao:2015,Zhen:2015} have attracted much attention; they describe open systems with loss (dissipation) and gain (e.g.\ coherent amplification in a laser)~\cite{Longhi:2017,El-Ganainy:2018}.
Extending topological methods to these systems may be particularly beneficial for the design of topological protected  laser modes~\cite{Bahari:2017,St-Jean:2017,Parto:2018}.
Moreover, \emph{genuinely non-Hermitian Hamiltonians}, i.e.\ Hamiltonians that cannot be deformed to a Hermitian Hamiltonian without energy bands touching, have novel topological properties not found in Hermitian systems.
They can be characterized by topological invariants different from those of Hermitian systems~\cite{Bergholtz:2019a,Gong:2018,Kawabata:2019b,Zhou:2019a,Shnerb:1998,Lee:2016,Leykam:2017,Shen:2018,Lieu:2018,Hirsbrunner:2019,Longhi:2019a,Chen:2018c}, but the extent of a bulk-boundary correspondence is, surprisingly, much less clear~\cite{Xiong:2018,Kunst:2018,Yao:2018b,Lee:2019c,Jin:2019a,Herviou:2019,Ge:2019,Borgnia:2020,Brzezicki:2019a,Yokomizo:2019,Kawabata:2020,Yang:2019g}.

We consider systems in one dimensions, which are particularly interesting because not only the eigenvectors but also the eigenenergies can have a nontrivial winding number.
In the case of a two-band model with chiral symmetry, the Bloch Hamiltonian is off-diagonal
%
\begin{equation}
    H(k) = \begin{pmatrix}
        0 & q_+(k) \\
        q_-(k) & 0
    \end{pmatrix}
\label{eq-bloch}
.\end{equation}
%
In a lattice model, the lattice spacing forces the momentum $k$ to be periodic, and the $q_{\pm}(k)$ describe closed paths in the complex plane. For example, a non-Hermitian Su-Schrieffer-Heeger (SSH) model is given by $q_{\pm}(k) = (m-1)  + e^{\mp i(k- i \gamma)}$, and the paths are circles with different radii centered on the real axis.~\setcounter{footnote}{28}\footnote{See Supplemental Material at \dots{} for non-Hermitian extensions of the SSH lattice model and an elementary discussion of the Mahaux-Weidenm{\"u}ller formula.}
The eigenvalues of the matrix $H(k)$ are distinct if neither path passes through the origin; in this case, we can assign to each path a winding number around the origin.
These form the $\Z\times\Z$ topological invariant of a non-Hermitian Hamiltonian in symmetry class AIII~\cite{Gong:2018}.
Hermitian Hamiltonians are characterized by $q_+(k)=q_-(k)^*$, which forces both winding numbers to be opposites of each other; a single $\Z$-invariant remains~\cite{Chiu:2016a,Schnyder:2008,Kitaev:2009}.
Genuinely non-Hermitian phases appear whenever the two winding numbers are no longer opposites of each other~\cite{Gong:2018,Leykam:2017}.
In this case, the \emph{non-Hermitian winding number}, which is the winding number of the determinant $\det(H(k))$, is nonzero.

Is there a bulk-boundary correspondence for the non-Hermitian winding number?
To answer this, we focus on  response (Green) functions,  which  describe experimental observables for instance in a scattering setup.
We find that the bulk-boundary correspondence breaks down once the non-Hermitian winding number takes a non-trivial value: When the winding number changes from zero, the bulk response starts exhibiting exponential \emph{growth} in space, and since periodic systems cannot accommodate such spatial growth, they do not reflect the properties of systems with open boundaries.
In this Letter, we focus on the specific  example of non-Hermitian Dirac fermions to discuss the above physics, while a general proof is contained in the companion paper Ref.~\cite{Zirnstein:2020}. The growth of the bulk response is distinct from the so-called non-Hermitian skin effect~\cite{Xiong:2018,Yao:2018b,MartinezAlvarez:2018,MartinezAlvarez:2018a,Zhang:2020e,Okuma:2020,Okuma:2020a,Xiao:2020,Helbig:2020,Hofmann:2020,Yoshida:2020,Longhi:2019d}.

\section{Example: Dirac fermions with non-Hermitian terms}
We consider a continuum model that corresponds to the long distance limit of the non-Hermitian SSH model~\cite{Lee:2016,Yao:2018b,Lieu:2018a,Esaki:2011}.
It concerns wave functions with two components $\psi (x)=[\psi _1(x),\psi _2(x)]^T$ subject to a Hermitian Dirac Hamiltonian 
$
H_0 = m\sigma _x + (-i\partial _x) \sigma _y
$,
where $\sigma _x,\sigma _y$ are Pauli matrices, $m$ is a real mass parameter (band gap).
Let us introduce non-Hermiticity by adding constant antihermitian terms:
\begin{equation}
    \hat H
    = \hat H_0 + i\gamma \sigma_y
\label{eq-model}
,\end{equation}
where $\gamma$ is real.
There are three more terms that we could add: $i\gamma_x \sigma_x$, $i\gamma_z \sigma_z$, and $-i\Gamma\Id$, where $\Id$ is the identity matrix.
The first can be absorbed by analytic continuation of the mass $m$.
The second and third vanish if we also impose a chiral symmetry, $\{\hat H,\sigma _z\}=0$, necessary for discussing zero energy boundary eigenstates in one dimension.
Thus, the symmetry class is AIII~\cite{Hasan:2010a} for complex $m$. For real mass $m$, $\hat H$ is additionally invariant under complex conjugation, placing it in symmetry class BDI, which also implies that eigenvalues occur in complex conjugate pairs.

\begin{figure}
\includegraphics[width=0.38\textwidth]{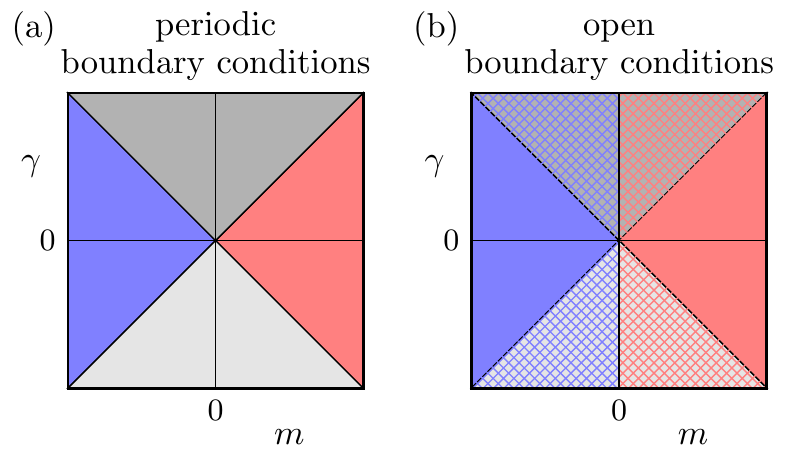}%
\caption{%
Topological phase diagram of one-dimensional non-Hermitian Dirac fermions with particle-hole symmetry and chiral symmetry.
(a) Periodic boundary conditions. Two winding numbers distinguish four phases: Two Hermitian (red and blue) and two genuinely non-Hermitian phases (grey), separated by lines $\gamma =\pm m$.
(b) Open boundary conditions.
The line $m=0$ separates phases with a different number of zero energy boundary eigenstates. For the boundary conditions \eqref{eq-boundary}, a positive mass implies the existence  of a boundary state at each end (red), which are absent for a  negative mass (blue).
The lines $\gamma =\pm m$ now indicate that the bulk (and boundary) Green function change from exponential decay to exponential growth.
\label{fig-phases}}
\end{figure}

In the continuum model~\eqref{eq-model}, we have $q_{\pm}(k)=m\pm(\gamma-ik)$, and the paths described by $q_{\pm}(k)$ in the complex plane are no longer closed. 
Still, one can assign a half-integer winding number~\cite{Leykam:2017} that changes whenever a path crosses the origin. Such crossings happen at $\gamma =\pm m$ and we find the topological phase diagram in Fig.~\ref{fig-phases}(a).

%

\subsection{Open boundary conditions}
We now consider a system of length $L$ with open boundary conditions
\begin{equation}
    \psi _2(0) = 0
    ,\ 
    \psi _1(L) = 0
\label{eq-boundary}
\end{equation}
corresponding to a particular boundary termination of the lattice model.
For open boundary conditions, 
the non-Hermitian terms in both the Dirac-Hamiltonian Eq.~\eqref{eq-model} and the non-Hermitian SSH model defined below Eq.~\eqref{eq-bloch}
can be eliminated by a similarity transformation: if $\psi _0(x)$ is an eigenfunction of the Hamiltonian $\hat H_0$, then $\psi (x)=e^{\gamma x}\psi _0(x)$ is an eigenfunction of the Hamiltonian $\hat H$. 
From this we see that \emph{all} eigenfunctions are exponentially localized.
This is the non-Hermitian skin effect~\cite{Xiong:2018,Yao:2018b,MartinezAlvarez:2018,MartinezAlvarez:2018a,Zhang:2020e,Okuma:2020,Okuma:2020a,Xiao:2020,Helbig:2020,Hofmann:2020,Yoshida:2020,Longhi:2019d}.

\subsection{Bulk and boundary Green function}
\label{section-green-boundary}
To clearly distinguish bulk and boundary, we now focus on Green functions, which are 
matrix-valued solutions to the equation 
\begin{equation}
    (E - \hat H)G(E;x,y) = \Id\delta(x-y)
\label{eq-green-differential}
\ \  .\end{equation}
The \emph{bulk Green function} $G_{\text{bulk}}$ is defined as the response of an infinite system~\cite{Zirnstein:2020},
whereas the Green function $G_{\text{open}}$ for open boundary conditions is defined as the solution that satisfies the conditions~\eqref{eq-boundary}.
When we probe the system far away from the boundary, $0 \ll x,y \ll L$, then only the bulk of the system responds, and we expect that both Green functions give the same result. However, when the source is close to the boundary, $y \approx 0$ or $y\approx L$, we expect that reflection at the boundary is important, which is captured in the \emph{boundary Green function}
\begin{equation}
    G_{\text{bound}}(E;x,y) := G_{\text{open}}(E;x,y) - G_{\text{bulk}}(E;x-y)
\label{eq-green-decomposition}
.\end{equation}
It solves the homogeneous equation $(E-\hat H)G_{\text{bound}}(E;x,y)=0$.
We have used that for a translationally invariant Hamiltonian, the bulk response only depends on the difference $x-y$.
If $G_0$ denotes a Green function of $\hat H_0$ for open boundaries, then the corresponding retarded Green function for $\hat H$ reads
\begin{equation}
    G(E;x,y) = G_0(E+i\eta;x,y)e^{\gamma (x-y)}
\label{eq-green-gauge}
,\end{equation}
with $\eta=0^+$.
We now focus on zero energy, $E=0$. Then, we find 
\begin{equation}
    G_{0,\text{bulk}}(i\eta;x,y) =
    [\theta (-\tilde{x})G_L + \theta (\tilde{x})G_R]
    e^{-\sqrt{m^2+\eta^2} |\tilde{x}|}
\label{eq-green-bulk-dirac}
,\end{equation}
where $\tilde{x}=x-y$, and $G_L$ and $G_R$ are matrices
\begin{equation}
    G_{s} = \mathcal{N}\begin{pmatrix}
        i\eta & m + \nu_s \sqrt{m^2+\eta ^2} \\
        m - \nu_s \sqrt{m^2+\eta ^2} & i\eta
    \end{pmatrix}
\end{equation}
with $s=L,R$, $\nu_{R/L} =\pm 1$, and $\mathcal{N}=1/(2\sqrt{m^2+\eta^2})$.
Thus, we obtain one of our main results: In the phases where the non-Hermitian winding number is nonzero, $|\gamma|>|m|$, the bulk Green function $G_{\text{bulk}}$ grows exponentially as $x\to \pm \infty$ while keeping $y$ fixed.
For $G_{0,\text{bound}}$ near the left boundary, we find
\begin{equation}
    G_{0,\text{bound}}(i\eta,x,y) = G_B e^{-\sqrt{m^2+\eta^2}(x+y)}
    ,\text{ for } x,y  \ll L
\label{eq-ssh-green-boundary}
.\end{equation}
Here, $G_B$ is the matrix
\begin{equation}
    G_B = -G_R\cdot \begin{pmatrix}
        \frac{m+\sqrt{m^2+\eta^2}}{m-\sqrt{m^2+\eta^2}} & 0 \\
        0 & 1
    \end{pmatrix}
.\end{equation}
Taken together, this yields the decomposition~\eqref{eq-green-decomposition}.

\subsection{Boundary eigenstates}
\label{section-boundary-green}
The Green function can be expressed as a sum over eigenstates
\begin{equation}
    G(E;x,y) = \sum_{n} (E-E_n)^{-1}\<x|\psi ^n_R\>\<\psi ^n_L|y\>
.\end{equation}
Here, $|\psi ^n_R\>$ are the so-called right- and $|\psi ^n_L\>$ the left eigenstates of the non-Hermitian Hamiltonian, i.e.~$H|\psi ^n_R\>=E_n|\psi ^n_R\>$ and $H^\dagger |\psi ^n_L\>=E_n^*|\psi ^n_L\>$~\cite{Brody:2014}.
The contribution $\<x|\psi ^n_R\>\<\psi ^n_L|y\>$ of an individual eigenstate to the Green function can be extracted as the residue of the pole at $E=E_n$~\cite{Peierls:1959,Zworski:1999}.
For identical positions $x=y$, this residue yields the biorthogonal polarization discussed in Ref.~\cite{Kunst:2018}.
We now define a \emph{boundary eigenstate} to be the residue of a pole of the \emph{boundary} Green function, and focus on states at zero energy, $E=0$.
For open boundary conditions, our model has only real eigenvalues due to the relation Eq.~(\ref{eq-green-gauge}), and we can obtain the residue from the imaginary part of the Green function  since $\Im G(E+i0^+;x,y) = -\sum_n \<x|\psi _R\>\<\psi _L|y\>\delta(E-E_n)$ for real $E$.
We find that 
\begin{align}
    - \Im G^{11}&_{\text{bound}}(0,x,y)
    = Ae^{(\gamma -m)x}e^{(-\gamma -m)y}
,\nonumber \\
    &\text{ where } A = \theta(m)2m/\eta \text{ with } \eta=0^+
\label{eq-ssh-state-boundary}
.\end{align}
Thus, for $m>0$, the boundary Green function has an isolated pole at zero energy, whose associated eigenstate is $\<x|\psi _R^0\> = e^{(\gamma -m)x}$ and $\<\psi _L^0|y\> = e^{(-\gamma -m)y}$.
The spatial shape changes dramatically from exponentially localized to exponentially growing and vice versa whenever $\gamma =\pm m$.
In contrast, for $m<0$, no boundary eigenstate is found.
Thus, the number of zero energy boundary eigenstates does not change during the topological phase transition at $\gamma = \pm |m|$ for periodic boundary conditions [Fig.~\ref{fig-phases}(b)], and the bulk-boundary correspondence breaks down. 

\begin{figure}
\includegraphics{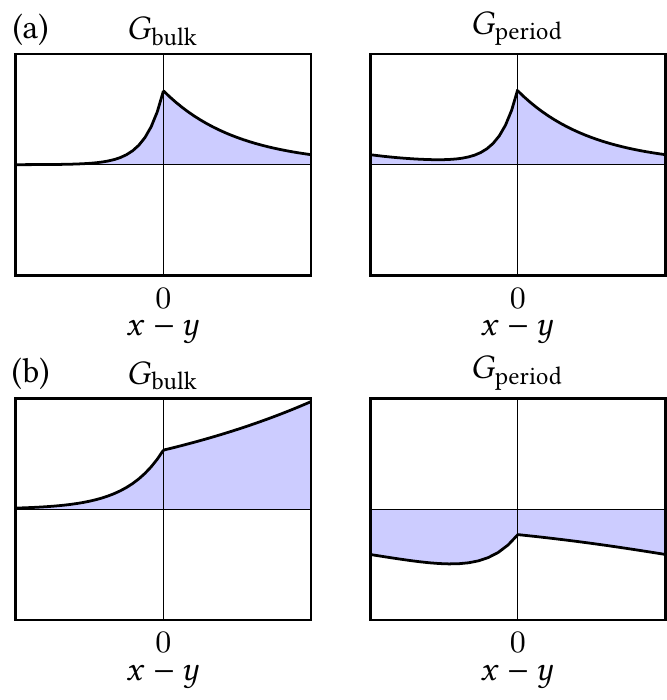}%
\caption{%
Breakdown of the bulk-periodic correspondence.
(a) If the bulk Green function decays spatially, then both bulk and periodic Green function agree.
(b) If the bulk Green function grows spatially, then the periodic Green function has to change drastically in order to accommodate periodic boundary conditions.
\label{fig-green}}%
\end{figure}

\subsection{Bulk-periodic correspondence}
\label{section-bulk-periodic}
The traditional view on the bulk-boundary correspondence actually comprises two separate logical steps: it relates i) the bulk to the boundary Green function, and ii) the Green function $G_{\text{period}}$ for periodic boundary conditions to that for the bulk of an infinite system: In the limit of large system size, both agree if the bulk Green function decays spatially~[Fig.~\ref{fig-green}(a)];
this allows us to use topological invariants of the Bloch Hamiltonian \eqref{eq-bloch} to characterize an infinite bulk.
In non-Hermitian systems, step i) is unproblematic, but step ii) may fail.
To better distinguish them, we propose to narrow the name \emph{bulk-boundary correspondence} to refer only to the first step, and to call the second step the \emph{bulk-periodic correspondence}.

Indeed, for our model in the regime $|\gamma |>|m|$, the bulk-periodic correspondence breaks down, because the periodic Green function decays, while the bulk Green function grows exponentially.~[Fig.~\ref{fig-green}(b)]
This growth also explains the exponential sensitivity to small perturbations seen in Ref.~\cite{Herviou:2019}.
For periodic boundary conditions $\psi (-L/2) = \psi (+L/2)$, and using the results Eqs.~\eqref{eq-green-gauge} and \eqref{eq-green-bulk-dirac} 
for the bulk Green function, we find 
\begin{align}
G_{\text{period}}(0;x,0)
    &= G_{\text{bulk}}(0;x,0)
\nonumber\\ &
    + G_L \frac{e^{\kappa _L x}}{e^{\kappa _L L}-1}
    + G_R \frac{e^{\kappa _R x}}{e^{-\kappa _R L}-1}
.\end{align}
with $\kappa _{L/R} = \gamma \pm \sqrt{m^2 + \eta^2}$.
In the limit of large system size, $L \gg |x|$, the two additional terms vanish if only if the exponents satisfy $\kappa _L > 0$ and $\kappa _R < 0$, i.e.\ if the bulk Green function decays spatially [Fig.~\ref{fig-green}].

If we focus on bulk growth and disregard boundary eigenstates, we no longer require symmetry class AIII. Then, we find our main result, which holds both with and without symmetry (class A): If the non-Hermitian winding number is nonzero, then the bulk Green function at zero energy grows spatially.
For example, consider a general Dirac model $\hat H = (-i\partial _x) \tau_1 + m_1 \tau_1 + m_2 \tau_2 + \dots + m_n \tau_n - i\Gamma\Id$ where the $\tau_j$ are Hermitian gamma matrices, $\{\tau_i,\tau_j\}=2\delta_{ij}$, the masses $m_j$ are complex and $\Gamma \geq 0$.
Then, since $(\hat H+i\Gamma)^2$ is proportional to the identity matrix, the corresponding bulk Green function  has form
$G_{\text{bulk}}(0;x,0) = G_L \theta (-x)e^{\kappa _L x} + G_R \theta (x) e^{\kappa _R x}$with $\kappa_{L,R} = -i m_1 \pm \sqrt{m_2^2 + \dots + m_n^2 + \Gamma^2}$ and appropriate $G_{L,R}$.
Here, the branch of the complex square root is the one with positive real part; we choose it by demanding that we remain in the same branch when $\Gamma\to\infty$~\cite{Zirnstein:2020}.
Thus, the Green function grows in space if and only if the imaginary part of $m_1$ exceeds the real part of the root.
Imposing symmetries will only constrain the parameters, but not affect this conclusion.
In general, the Green function is a sum of exponentials $\exp(ik_s x)$ where the complex momenta $k_s$ are the zeros of $\det(H(k))$; the subscript $s$ refers to a side $L,R$ and an index.
For local continuum models like the Dirac model, this determinant is a polynomial in momentum $k$.
Thus, the non-Hermitian winding number $\nu(H) = (2\pi)^{-1}\int_{-\infty}^{\infty} dk\, \partial_k \arg\det(H(k))$ is the sum of $+1/2$ for each zero above the real axis and $-1/2$ for each zero below.
But this number changes precisely when one of the zeros crosses the real axis, which means that the exponential changes from spatial decay to spatial growth.
We extend this sketch to a full proof for lattice models in Ref.~\cite{Zirnstein:2020}.
While for the above Dirac models, exponential growth only occurs in the genuinely non-Hermitian phases, for other models, it may arise even when the Hamiltonian can be deformed to a Hermitian one; see~\cite{Note29} for an example. Thus, the growth of the bulk Green function is not, by itself, topologically invariant.

Our work still leaves open the exciting question of the bulk-boundary correspondence in the narrow sense: Are there topological invariants of the bulk Green function that imply the presence of  boundary eigenstates? For Dirac fermions, the latter persist well into the genuinely non-Hermitian phases. This is also true for lattice models discussed in the literature~\cite{Lee:2016,Kunst:2018,Yao:2018b,Xiong:2018,Lee:2019c,Herviou:2019}, see the Supplemental Material~\cite{Note29} for details.

\section{Experimental response: Scattering}
\label{section-scattering}
%
\begin{figure}
\includegraphics[width=0.48\textwidth]{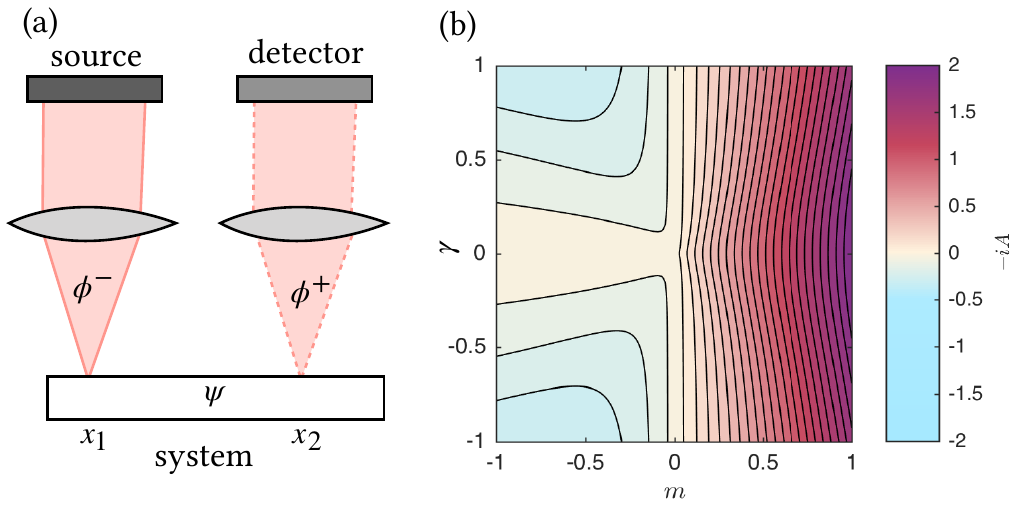}%
\caption{%
(a) Scattering setup.
(b) Scattering response of the one-dimensional Dirac fermion with dissipation $\Gamma =|\gamma|$.
The plot shows the amplitude $-iA$ of the boundary response at the left boundary.%
\label{fig-scattering}}
\end{figure}
In a scattering configuration [see Fig.~\ref{fig-scattering}(a)], an excitation of an outside field $\phi$ is created, and the incoming amplitude $\phi^-$ targets a point $x_1$ of the system.
If the excitation is monochromatic with frequency (energy) $E$, then the system will eventually reach a stationary state $\psi$ that, in turn, emits an outgoing amplitude $\phi ^+$ at every position $x_2$.
The scattering matrix $S(E)$ records how incoming amplitudes are mapped to outgoing amplitudes, $\phi ^+ = S(E)\phi ^-$,  
and is given by the Mahaux-Weidenm\"{u}ller formula~\cite{Mahaux:1969}
\begin{align}
    S(E) &= \Id - 2i W^\dagger \frac1{E - \hat H} W
    ,\quad
    \hat H = \hat H_0 - iWW^\dagger 
\label{eq-scattering-matrix}
\end{align}
Here, the matrix $W$ describes how the outside field couples into the system. The Hamiltonian $\hat H_0$ describes the time evolution of the system if the coupling was absent; it is usually Hermitian, for it is the combination with the dissipative term $-iWW^\dagger $ that yields an effective non-Hermitian Hamiltonian $\hat H$.
The scattering matrix differs from unity by $-2iW^\dagger G(E)W$ where $G(E)$ is the Green function of this non-Hermitian Hamiltonian. 
To make this Letter self-contained, in the Supplemental Material~\cite{Note29} we include an elementary discussion of the Mahaux-Weidenm\"{u}ller formula ~\cite{Livsic:1973,Fyodorov:2000}, which is equivalent to \emph{temporal coupled mode-theory}~\cite{Suh:2004,Fan:2003} in optics.

So far, we have shown how coupling to an environment yields a non-Hermitian Hamiltonian.
Conversely, we now consider a Hamiltonian $H$ and attempt to realize it in a scattering setup.
For this, we decompose it as $\hat H=\hat H_0 + i\hat \Gamma $ where $\hat H_0=[\hat H + \hat H^\dagger ]/2$ and $\hat \Gamma =(i/2)[\hat{H}^\dagger -\hat H]$ are Hermitian matrices. 
We define a Hamiltonian to be \emph{purely dissipative} if $\hat \Gamma $ is negative semidefinite, $\hat \Gamma \leq 0$, i.e.\ if it has no positive eigenvalues.
Any matrix of the form $\hat \Gamma =-WW^\dagger $ is negative semidefinite, and any semidefinite $\hat \Gamma $ can be represented in the above  form by choosing $W=(-\hat \Gamma )^{1/2}$.
This choice is unique up to a change of basis for the outside field, $W \to WU$, with $U$ unitary, and up to components that do not couple.
Thus, any purely dissipative Hamiltonian can be realized in a  scattering setup described by Eq.~\eqref{eq-scattering-matrix}. 
But in order to realize a general non-Hermitian Hamiltonian, we have to allow for positive eigenvalues in the antihermitian part, which corresponds to (linear) \emph{gain}, i.e.\ coherent amplification.
In fact, gain is a key requirement for the non-Hermitian winding number to be nonzero, see Ref.~\cite{Zirnstein:2020}.

In optical systems, gain can be realized by optical pumping~\footnote{Exponential amplification cannot be indefinite; in practice, gain will saturate nonlinearly~\cite{Malzard:2018a}}, and spatially uniform gain can be described by the ansatz $\hat H = \hat H_0 + ig - iWW^\dagger $ with a positive constant $g$ in Eq.~\eqref{eq-scattering-matrix}.
Then, the antihermitian part of $\hat H$ is related to the coupling matrix via $- WW^\dagger = \hat \Gamma - g$, and this relation can be solved for $W$ if the non-Hermitian Hamiltonian $\hat H$ has \emph{bounded gain}, i.e.\ if $\hat \Gamma - g$  becomes negative semidefinite by making $g$ sufficiently large.
For example, the Dirac fermion, Eq.~\eqref{eq-model}, can be described by 
$\hat H = \hat H_0 + ig - iWW^\dagger $ with $g = |\gamma|$, $W = \sqrt{|\gamma|} [1, -\sgn(\gamma )i]^T$;
the coupling $W$ is a $2\times 1$ matrix when modeling the outside field as a scalar.

Since a purely dissipative setup can be realized more easily than one with gain, we ask whether signatures of the zero energy boundary states of the Dirac fermion Eq.~(\ref{eq-model}) can still be found without gain.
This corresponds to allowing a constant dissipation, $\hat H = \hat H_0 + i\gamma \sigma_y - i\Gamma \Id$, which does break chiral symmetry.
We consider the special case $\Gamma =|\gamma|$, which has the least amount of dissipation while probing the phase transition region
$\gamma = \pm |m|$.
At zero energy, we can use Eq.~\eqref{eq-ssh-green-boundary} with $\eta=\Gamma$ and find
\begin{align}
    [W^\dagger G_{\text{bound}}(&E=0) W](x_2,x_1=0)
    =
    A\exp\left(-x_2/\xi _L\right)
,\nonumber \\
    A &= -i \frac{m}{\sqrt{m^2+\gamma ^2}} \frac{\gamma ^2}{m-\sqrt{m^2+\gamma ^2}}
,\end{align}
where $A$ denotes the amplitude and $\xi _L=1/(\sqrt{m^2+\gamma ^2}-\gamma )$ the localization length of the response.
We see that the amplitude can still distinguish the phase with boundary eigenstate, $m>0$, from the phase without, $m<0$, though the distinction is less sharp [Fig.~\ref{fig-scattering}(b)] than in the case with gain and chiral symmetry.

\section{Conclusion}
Using the example of non-Hermitian Dirac fermions, we have discussed the topological phase diagram via Green functions. The key idea was to decompose the Green function for open boundary conditions into a bulk and a boundary Green function, Eq.~\eqref{eq-green-decomposition}.
We have shown that a nonzero non-Hermitian winding number means that the correspondence of the open system to a  periodic one breaks down due to an exponential growth of the bulk Green function.
We have also discussed how to realize non-Hermitian Hamiltonian in a scattering setup and indicated that observing a nonzero topological invariant requires a system with gain.

\section{Acknowledgements}
We would like to thank T.~Karzig for helpful discussions. 
B.~R.\ and H.-G.~Z.\ acknowledge financial support from the German Research Foundation within the Collaborative Research Centre 762 (project B6).
B.~R.\ acknowledges support from the Rosi and Max Varon Visiting Professorship at the Weizmann Institute of Science. We are grateful for the hospitality of the Aspen Center for Physics, funded by NSF grant PHY-1607611, where part of this work was performed.
G.~R.\ is grateful for generous support from the Institute of Quantum Information and Matter, an NSF frontier center, NSF grant 1839271, and The Simons Foundation.

\bibliography{bib-non-hermitian}

\begin{thebibliography}{76}%
\makeatletter
\providecommand \@ifxundefined [1]{%
 \@ifx{#1\undefined}
}%
\providecommand \@ifnum [1]{%
 \ifnum #1\expandafter \@firstoftwo
 \else \expandafter \@secondoftwo
 \fi
}%
\providecommand \@ifx [1]{%
 \ifx #1\expandafter \@firstoftwo
 \else \expandafter \@secondoftwo
 \fi
}%
\providecommand \natexlab [1]{#1}%
\providecommand \enquote  [1]{``#1''}%
\providecommand \bibnamefont  [1]{#1}%
\providecommand \bibfnamefont [1]{#1}%
\providecommand \citenamefont [1]{#1}%
\providecommand \href@noop [0]{\@secondoftwo}%
\providecommand \href [0]{\begingroup \@sanitize@url \@href}%
\providecommand \@href[1]{\@@startlink{#1}\@@href}%
\providecommand \@@href[1]{\endgroup#1\@@endlink}%
\providecommand \@sanitize@url [0]{\catcode `\\12\catcode `\$12\catcode
  `\&12\catcode `\#12\catcode `\^12\catcode `\_12\catcode `\%12\relax}%
\providecommand \@@startlink[1]{}%
\providecommand \@@endlink[0]{}%
\providecommand \url  [0]{\begingroup\@sanitize@url \@url }%
\providecommand \@url [1]{\endgroup\@href {#1}{\urlprefix }}%
\providecommand \urlprefix  [0]{URL }%
\providecommand \Eprint [0]{\href }%
\providecommand \doibase [0]{http://dx.doi.org/}%
\providecommand \selectlanguage [0]{\@gobble}%
\providecommand \bibinfo  [0]{\@secondoftwo}%
\providecommand \bibfield  [0]{\@secondoftwo}%
\providecommand \translation [1]{[#1]}%
\providecommand \BibitemOpen [0]{}%
\providecommand \bibitemStop [0]{}%
\providecommand \bibitemNoStop [0]{.\EOS\space}%
\providecommand \EOS [0]{\spacefactor3000\relax}%
\providecommand \BibitemShut  [1]{\csname bibitem#1\endcsname}%
\let\auto@bib@innerbib\@empty
\bibitem [{\citenamefont {Hasan}\ and\ \citenamefont
  {Kane}(2010)}]{Hasan:2010a}%
  \BibitemOpen
  \bibfield  {author} {\bibinfo {author} {\bibfnamefont {M~Zahid}\ \bibnamefont
  {Hasan}}\ and\ \bibinfo {author} {\bibfnamefont {C~L}\ \bibnamefont {Kane}},\
  }\bibfield  {title} {{\selectlanguage {English}\enquote {\bibinfo {title}
  {Colloquium: {{Topological}} insulators},}\ }}\href {\doibase
  10.1103/RevModPhys.82.3045} {\bibfield  {journal} {\bibinfo  {journal} {Rev.
  Mod. Phys.}\ }\textbf {\bibinfo {volume} {82}},\ \bibinfo {pages}
  {3045--3067} (\bibinfo {year} {2010})}\BibitemShut {NoStop}%
\bibitem [{\citenamefont {Qi}\ and\ \citenamefont {Zhang}(2011)}]{Qi:2011}%
  \BibitemOpen
  \bibfield  {author} {\bibinfo {author} {\bibfnamefont {Xiao-Liang}\
  \bibnamefont {Qi}}\ and\ \bibinfo {author} {\bibfnamefont {Shou-Cheng}\
  \bibnamefont {Zhang}},\ }\bibfield  {title} {{\selectlanguage
  {English}\enquote {\bibinfo {title} {Topological insulators and
  superconductors},}\ }}\href {\doibase 10.1103/RevModPhys.83.1057} {\bibfield
  {journal} {\bibinfo  {journal} {Rev. Mod. Phys.}\ }\textbf {\bibinfo {volume}
  {83}},\ \bibinfo {pages} {1057--1110} (\bibinfo {year} {2011})}\BibitemShut
  {NoStop}%
\bibitem [{\citenamefont {Chiu}\ \emph {et~al.}(2016)\citenamefont {Chiu},
  \citenamefont {Teo}, \citenamefont {Schnyder},\ and\ \citenamefont
  {Ryu}}]{Chiu:2016a}%
  \BibitemOpen
  \bibfield  {author} {\bibinfo {author} {\bibfnamefont {Ching-Kai}\
  \bibnamefont {Chiu}}, \bibinfo {author} {\bibfnamefont {Jeffrey C~Y}\
  \bibnamefont {Teo}}, \bibinfo {author} {\bibfnamefont {Andreas~P}\
  \bibnamefont {Schnyder}}, \ and\ \bibinfo {author} {\bibfnamefont {Shinsei}\
  \bibnamefont {Ryu}},\ }\bibfield  {title} {\enquote {\bibinfo {title}
  {Classification of topological quantum matter with symmetries},}\ }\href
  {\doibase 10.1103/RevModPhys.88.035005} {\bibfield  {journal} {\bibinfo
  {journal} {Rev. Mod. Phys.}\ }\textbf {\bibinfo {volume} {88}},\ \bibinfo
  {pages} {035005} (\bibinfo {year} {2016})},\ \Eprint
  {http://arxiv.org/abs/1505.03535v2} {arXiv:1505.03535v2} \BibitemShut
  {NoStop}%
\bibitem [{\citenamefont {Bernevig}\ and\ \citenamefont
  {Hughes}(2013)}]{Bernevig:2013}%
  \BibitemOpen
  \bibfield  {author} {\bibinfo {author} {\bibfnamefont {B~Andrei}\
  \bibnamefont {Bernevig}}\ and\ \bibinfo {author} {\bibfnamefont {Taylor~L}\
  \bibnamefont {Hughes}},\ }\href@noop {} {\emph {\bibinfo {title} {Topological
  Insulators and Topological Superconductors}}}\ (\bibinfo  {publisher}
  {{Princeton University Press}},\ \bibinfo {year} {2013})\BibitemShut
  {NoStop}%
\bibitem [{\citenamefont {Bansil}\ \emph {et~al.}(2016)\citenamefont {Bansil},
  \citenamefont {Lin},\ and\ \citenamefont {Das}}]{Bansil:2016}%
  \BibitemOpen
  \bibfield  {author} {\bibinfo {author} {\bibfnamefont {A}~\bibnamefont
  {Bansil}}, \bibinfo {author} {\bibfnamefont {Hsin}\ \bibnamefont {Lin}}, \
  and\ \bibinfo {author} {\bibfnamefont {Tanmoy}\ \bibnamefont {Das}},\
  }\bibfield  {title} {{\selectlanguage {English}\enquote {\bibinfo {title}
  {Colloquium: {{Topological}} band theory},}\ }}\href {\doibase
  10.1103/RevModPhys.88.021004} {\bibfield  {journal} {\bibinfo  {journal}
  {Rev. Mod. Phys.}\ }\textbf {\bibinfo {volume} {88}},\ \bibinfo {pages}
  {021004} (\bibinfo {year} {2016})}\BibitemShut {NoStop}%
\bibitem [{\citenamefont {Raghu}\ and\ \citenamefont
  {Haldane}(2008)}]{Raghu:2008}%
  \BibitemOpen
  \bibfield  {author} {\bibinfo {author} {\bibfnamefont {S}~\bibnamefont
  {Raghu}}\ and\ \bibinfo {author} {\bibfnamefont {F~D~M}\ \bibnamefont
  {Haldane}},\ }\bibfield  {title} {{\selectlanguage {English}\enquote
  {\bibinfo {title} {Analogs of quantum-{{Hall}}-effect edge states in photonic
  crystals},}\ }}\href {\doibase 10.1103/PhysRevA.78.033834} {\bibfield
  {journal} {\bibinfo  {journal} {Phys. Rev. A}\ }\textbf {\bibinfo {volume}
  {78}},\ \bibinfo {pages} {033834} (\bibinfo {year} {2008})}\BibitemShut
  {NoStop}%
\bibitem [{\citenamefont {Lu}\ \emph {et~al.}(2014)\citenamefont {Lu},
  \citenamefont {Joannopoulos},\ and\ \citenamefont {Solja{\v
  c}i{\'c}}}]{Lu:2014}%
  \BibitemOpen
  \bibfield  {author} {\bibinfo {author} {\bibfnamefont {Ling}\ \bibnamefont
  {Lu}}, \bibinfo {author} {\bibfnamefont {John~D.}\ \bibnamefont
  {Joannopoulos}}, \ and\ \bibinfo {author} {\bibfnamefont {Marin}\
  \bibnamefont {Solja{\v c}i{\'c}}},\ }\bibfield  {title} {{\selectlanguage
  {English}\enquote {\bibinfo {title} {Topological photonics},}\ }}\href {\doibase
  10.1038/nphoton.2014.248} {\bibfield  {journal} {\bibinfo  {journal} {Nat.
  Photonics}\ }\textbf {\bibinfo {volume} {8}},\ \bibinfo {pages} {821--829}
  (\bibinfo {year} {2014})}\BibitemShut {NoStop}%
\bibitem [{\citenamefont {Khanikaev}\ and\ \citenamefont
  {Shvets}(2017)}]{Khanikaev:2017}%
  \BibitemOpen
  \bibfield  {author} {\bibinfo {author} {\bibfnamefont {Alexander~B}\
  \bibnamefont {Khanikaev}}\ and\ \bibinfo {author} {\bibfnamefont {Gennady}\
  \bibnamefont {Shvets}},\ }\bibfield  {title} {{\selectlanguage
  {English}\enquote {\bibinfo {title} {Two-dimensional topological
  photonics},}\ }}\href {\doibase 10.1038/s41566-017-0048-5} {\bibfield
  {journal} {\bibinfo  {journal} {Nat. Photonics}\ }\textbf {\bibinfo {volume}
  {11}},\ \bibinfo {pages} {763--773} (\bibinfo {year} {2017})}\BibitemShut
  {NoStop}%
\bibitem [{\citenamefont {Rechtsman}\ \emph {et~al.}(2013)\citenamefont
  {Rechtsman}, \citenamefont {Zeuner}, \citenamefont {Plotnik}, \citenamefont
  {Lumer}, \citenamefont {Podolsky}, \citenamefont {Dreisow}, \citenamefont
  {Nolte}, \citenamefont {Segev},\ and\ \citenamefont
  {Szameit}}]{Rechtsman:2013}%
  \BibitemOpen
  \bibfield  {author} {\bibinfo {author} {\bibfnamefont {Mikael~C.}\
  \bibnamefont {Rechtsman}}, \bibinfo {author} {\bibfnamefont {Julia~M.}\
  \bibnamefont {Zeuner}}, \bibinfo {author} {\bibfnamefont {Yonatan}\
  \bibnamefont {Plotnik}}, \bibinfo {author} {\bibfnamefont {Yaakov}\
  \bibnamefont {Lumer}}, \bibinfo {author} {\bibfnamefont {Daniel}\
  \bibnamefont {Podolsky}}, \bibinfo {author} {\bibfnamefont {Felix}\
  \bibnamefont {Dreisow}}, \bibinfo {author} {\bibfnamefont {Stefan}\
  \bibnamefont {Nolte}}, \bibinfo {author} {\bibfnamefont {Mordechai}\
  \bibnamefont {Segev}}, \ and\ \bibinfo {author} {\bibfnamefont {Alexander}\
  \bibnamefont {Szameit}},\ }\bibfield  {title} {{\selectlanguage {English}\enquote
  {\bibinfo {title} {Photonic {{Floquet}} topological insulators},}\ }}\href
  {\doibase 10.1038/nature12066} {\bibfield  {journal} {\bibinfo  {journal}
  {Nature}\ }\textbf {\bibinfo {volume} {496}},\ \bibinfo {pages} {196--200}
  (\bibinfo {year} {2013})}\BibitemShut {NoStop}%
\bibitem [{\citenamefont {Plotnik}\ \emph {et~al.}(2014)\citenamefont
  {Plotnik}, \citenamefont {Rechtsman}, \citenamefont {Song}, \citenamefont
  {Heinrich}, \citenamefont {Zeuner}, \citenamefont {Nolte}, \citenamefont
  {Lumer}, \citenamefont {Malkova}, \citenamefont {Xu}, \citenamefont
  {Szameit}, \citenamefont {Chen},\ and\ \citenamefont {Segev}}]{Plotnik:2014}%
  \BibitemOpen
  \bibfield  {author} {\bibinfo {author} {\bibfnamefont {Yonatan}\ \bibnamefont
  {Plotnik}}, \bibinfo {author} {\bibfnamefont {Mikael~C.}\ \bibnamefont
  {Rechtsman}}, \bibinfo {author} {\bibfnamefont {Daohong}\ \bibnamefont
  {Song}}, \bibinfo {author} {\bibfnamefont {Matthias}\ \bibnamefont
  {Heinrich}}, \bibinfo {author} {\bibfnamefont {Julia~M.}\ \bibnamefont
  {Zeuner}}, \bibinfo {author} {\bibfnamefont {Stefan}\ \bibnamefont {Nolte}},
  \bibinfo {author} {\bibfnamefont {Yaakov}\ \bibnamefont {Lumer}}, \bibinfo
  {author} {\bibfnamefont {Natalia}\ \bibnamefont {Malkova}}, \bibinfo {author}
  {\bibfnamefont {Jingjun}\ \bibnamefont {Xu}}, \bibinfo {author}
  {\bibfnamefont {Alexander}\ \bibnamefont {Szameit}}, \bibinfo {author}
  {\bibfnamefont {Zhigang}\ \bibnamefont {Chen}}, \ and\ \bibinfo {author}
  {\bibfnamefont {Mordechai}\ \bibnamefont {Segev}},\ }\bibfield  {title}
  {{\selectlanguage {English}\enquote {\bibinfo {title} {Observation of
  unconventional edge states in `photonic graphene'},}\ }}\href {\doibase
  10.1038/nmat3783} {\bibfield  {journal} {\bibinfo  {journal} {Nat. Mater.}\
  }\textbf {\bibinfo {volume} {13}},\ \bibinfo {pages} {57--62} (\bibinfo
  {year} {2014})}\BibitemShut {NoStop}%
\bibitem [{\citenamefont {Noh}\ \emph {et~al.}(2018)\citenamefont {Noh},
  \citenamefont {Benalcazar}, \citenamefont {Huang}, \citenamefont {Collins},
  \citenamefont {Chen}, \citenamefont {Hughes},\ and\ \citenamefont
  {Rechtsman}}]{Noh:2018}%
  \BibitemOpen
  \bibfield  {author} {\bibinfo {author} {\bibfnamefont {Jiho}\ \bibnamefont
  {Noh}}, \bibinfo {author} {\bibfnamefont {Wladimir~A.}\ \bibnamefont
  {Benalcazar}}, \bibinfo {author} {\bibfnamefont {Sheng}\ \bibnamefont
  {Huang}}, \bibinfo {author} {\bibfnamefont {Matthew~J.}\ \bibnamefont
  {Collins}}, \bibinfo {author} {\bibfnamefont {Kevin~P.}\ \bibnamefont
  {Chen}}, \bibinfo {author} {\bibfnamefont {Taylor~L.}\ \bibnamefont
  {Hughes}}, \ and\ \bibinfo {author} {\bibfnamefont {Mikael~C.}\ \bibnamefont
  {Rechtsman}},\ }\bibfield  {title} {{\selectlanguage {English}\enquote {\bibinfo
  {title} {Topological protection of photonic mid-gap defect modes},}\ }}\href
  {\doibase 10.1038/s41566-018-0179-3} {\bibfield  {journal} {\bibinfo
  {journal} {Nat. Photonics}\ }\textbf {\bibinfo {volume} {12}},\ \bibinfo
  {pages} {408} (\bibinfo {year} {2018})}\BibitemShut {NoStop}%
\bibitem [{\citenamefont {Halperin}(1982)}]{Halperin:1982}%
  \BibitemOpen
  \bibfield  {author} {\bibinfo {author} {\bibfnamefont {B.~I.}\ \bibnamefont
  {Halperin}},\ }\bibfield  {title} {\enquote {\bibinfo {title} {Quantized
  {{Hall}} conductance, current-carrying edge states, and the existence of
  extended states in a two-dimensional disordered potential},}\ }\href
  {\doibase 10.1103/PhysRevB.25.2185} {\bibfield  {journal} {\bibinfo
  {journal} {Phys. Rev. B}\ }\textbf {\bibinfo {volume} {25}},\ \bibinfo
  {pages} {2185--2190} (\bibinfo {year} {1982})}\BibitemShut {NoStop}%
\bibitem [{\citenamefont {Hatsugai}(1993)}]{Hatsugai:1993}%
  \BibitemOpen
  \bibfield  {author} {\bibinfo {author} {\bibfnamefont {Yasuhiro}\
  \bibnamefont {Hatsugai}},\ }\bibfield  {title} {\enquote {\bibinfo {title}
  {Chern number and edge states in the integer quantum {{Hall}} effect},}\
  }\href {\doibase 10.1103/PhysRevLett.71.3697} {\bibfield  {journal} {\bibinfo
   {journal} {Phys. Rev. Lett.}\ }\textbf {\bibinfo {volume} {71}},\ \bibinfo
  {pages} {3697--3700} (\bibinfo {year} {1993})}\BibitemShut {NoStop}%
\bibitem [{\citenamefont {Essin}\ and\ \citenamefont
  {Gurarie}(2011)}]{Essin:2011}%
  \BibitemOpen
  \bibfield  {author} {\bibinfo {author} {\bibfnamefont {Andrew~M.}\
  \bibnamefont {Essin}}\ and\ \bibinfo {author} {\bibfnamefont {Victor}\
  \bibnamefont {Gurarie}},\ }\bibfield  {title} {{\selectlanguage {English}\enquote
  {\bibinfo {title} {Bulk-boundary correspondence of topological insulators
  from their respective {{Green}}'s functions},}\ }}\href {\doibase
  10.1103/PhysRevB.84.125132} {\bibfield  {journal} {\bibinfo  {journal} {Phys.
  Rev. B}\ }\textbf {\bibinfo {volume} {84}},\ \bibinfo {pages} {125132}
  (\bibinfo {year} {2011})}\BibitemShut {NoStop}%
\bibitem [{\citenamefont {Graf}\ and\ \citenamefont {Porta}(2013)}]{Graf:2013}%
  \BibitemOpen
  \bibfield  {author} {\bibinfo {author} {\bibfnamefont {Gian~Michele}\
  \bibnamefont {Graf}}\ and\ \bibinfo {author} {\bibfnamefont {Marcello}\
  \bibnamefont {Porta}},\ }\bibfield  {title} {{\selectlanguage {English}\enquote
  {\bibinfo {title} {Bulk-{{Edge Correspondence}} for {{Two}}-{{Dimensional
  Topological Insulators}}},}\ }}\href {\doibase 10.1007/s00220-013-1819-6}
  {\bibfield  {journal} {\bibinfo  {journal} {Commun. Math. Phys.}\ }\textbf
  {\bibinfo {volume} {324}},\ \bibinfo {pages} {851--895} (\bibinfo {year}
  {2013})}\BibitemShut {NoStop}%
\bibitem [{\citenamefont {Avila}\ \emph {et~al.}(2013)\citenamefont {Avila},
  \citenamefont {{Schulz-Baldes}},\ and\ \citenamefont
  {{Villegas-Blas}}}]{Avila:2013}%
  \BibitemOpen
  \bibfield  {author} {\bibinfo {author} {\bibfnamefont {Julio~Cesar}\
  \bibnamefont {Avila}}, \bibinfo {author} {\bibfnamefont {Hermann}\
  \bibnamefont {{Schulz-Baldes}}}, \ and\ \bibinfo {author} {\bibfnamefont
  {Carlos}\ \bibnamefont {{Villegas-Blas}}},\ }\bibfield  {title}
  {{\selectlanguage {English}\enquote {\bibinfo {title} {Topological {{Invariants}}
  of {{Edge States}} for {{Periodic Two}}-{{Dimensional Models}}},}\ }}\href
  {\doibase 10.1007/s11040-012-9123-9} {\bibfield  {journal} {\bibinfo
  {journal} {Math Phys Anal Geom}\ }\textbf {\bibinfo {volume} {16}},\ \bibinfo
  {pages} {137--170} (\bibinfo {year} {2013})}\BibitemShut {NoStop}%
\bibitem [{\citenamefont {Hatano}\ and\ \citenamefont
  {Nelson}(1996)}]{Hatano:1996}%
  \BibitemOpen
  \bibfield  {author} {\bibinfo {author} {\bibfnamefont {Naomichi}\
  \bibnamefont {Hatano}}\ and\ \bibinfo {author} {\bibfnamefont {David~R.}\
  \bibnamefont {Nelson}},\ }\bibfield  {title} {\enquote {\bibinfo {title}
  {Localization {{Transitions}} in {{Non}}-{{Hermitian Quantum Mechanics}}},}\
  }\href {\doibase 10.1103/PhysRevLett.77.570} {\bibfield  {journal} {\bibinfo
  {journal} {Phys. Rev. Lett.}\ }\textbf {\bibinfo {volume} {77}},\ \bibinfo
  {pages} {570--573} (\bibinfo {year} {1996})}\BibitemShut {NoStop}%
\bibitem [{\citenamefont {Bender}\ and\ \citenamefont
  {Boettcher}(1998)}]{Bender:1998}%
  \BibitemOpen
  \bibfield  {author} {\bibinfo {author} {\bibfnamefont {Carl~M}\ \bibnamefont
  {Bender}}\ and\ \bibinfo {author} {\bibfnamefont {Stefan}\ \bibnamefont
  {Boettcher}},\ }\bibfield  {title} {\enquote {\bibinfo {title} {Real
  {{Spectra}} in {{Non}}-{{Hermitian Hamiltonians Having PT Symmetry}}},}\
  }\href {\doibase 10.1103/PhysRevLett.80.5243} {\bibfield  {journal} {\bibinfo
   {journal} {Phys. Rev. Lett.}\ }\textbf {\bibinfo {volume} {80}},\ \bibinfo
  {pages} {5243--5246} (\bibinfo {year} {1998})}\BibitemShut {NoStop}%
\bibitem [{\citenamefont {Rotter}(2009)}]{Rotter:2009}%
  \BibitemOpen
  \bibfield  {author} {\bibinfo {author} {\bibfnamefont {Ingrid}\ \bibnamefont
  {Rotter}},\ }\bibfield  {title} {\enquote {\bibinfo {title} {A
  non-{{Hermitian Hamilton}} operator and the physics of open quantum
  systems},}\ }\href {\doibase 10.1088/1751-8113/42/15/153001} {\bibfield
  {journal} {\bibinfo  {journal} {J. Phys. Math. Theor.}\ }\textbf {\bibinfo
  {volume} {42}},\ \bibinfo {pages} {153001} (\bibinfo {year}
  {2009})}\BibitemShut {NoStop}%
\bibitem [{\citenamefont {Cao}\ and\ \citenamefont {Wiersig}(2015)}]{Cao:2015}%
  \BibitemOpen
  \bibfield  {author} {\bibinfo {author} {\bibfnamefont {Hui}\ \bibnamefont
  {Cao}}\ and\ \bibinfo {author} {\bibfnamefont {Jan}\ \bibnamefont
  {Wiersig}},\ }\bibfield  {title} {{\selectlanguage {English}\enquote
  {\bibinfo {title} {Dielectric microcavities: {{Model}} systems for wave chaos
  and non-{{Hermitian}} physics},}\ }}\href {\doibase 10.1103/RevModPhys.87.61}
  {\bibfield  {journal} {\bibinfo  {journal} {Rev. Mod. Phys.}\ }\textbf
  {\bibinfo {volume} {87}},\ \bibinfo {pages} {61--111} (\bibinfo {year}
  {2015})}\BibitemShut {NoStop}%
\bibitem [{\citenamefont {Zhen}\ \emph {et~al.}(2015)\citenamefont {Zhen},
  \citenamefont {Hsu}, \citenamefont {Igarashi}, \citenamefont {Lu},
  \citenamefont {Kaminer}, \citenamefont {Pick}, \citenamefont {Chua},
  \citenamefont {Joannopoulos},\ and\ \citenamefont {Solja{\v
  c}ic}}]{Zhen:2015}%
  \BibitemOpen
  \bibfield  {author} {\bibinfo {author} {\bibfnamefont {Bo}~\bibnamefont
  {Zhen}}, \bibinfo {author} {\bibfnamefont {Chia~Wei}\ \bibnamefont {Hsu}},
  \bibinfo {author} {\bibfnamefont {Yuichi}\ \bibnamefont {Igarashi}}, \bibinfo
  {author} {\bibfnamefont {Ling}\ \bibnamefont {Lu}}, \bibinfo {author}
  {\bibfnamefont {Ido}\ \bibnamefont {Kaminer}}, \bibinfo {author}
  {\bibfnamefont {Adi}\ \bibnamefont {Pick}}, \bibinfo {author} {\bibfnamefont
  {Song-Liang}\ \bibnamefont {Chua}}, \bibinfo {author} {\bibfnamefont
  {John~D}\ \bibnamefont {Joannopoulos}}, \ and\ \bibinfo {author}
  {\bibfnamefont {Marin}\ \bibnamefont {Solja{\v c}ic}},\ }\bibfield  {title}
  {\enquote {\bibinfo {title} {Spawning rings of exceptional points out of
  {{Dirac}} cones},}\ }\href {\doibase 10.1038/nature14889} {\bibfield
  {journal} {\bibinfo  {journal} {Nature}\ }\textbf {\bibinfo {volume} {525}},\
  \bibinfo {pages} {354--358} (\bibinfo {year} {2015})}\BibitemShut {NoStop}%
\bibitem [{\citenamefont {Longhi}(2017)}]{Longhi:2017}%
  \BibitemOpen
  \bibfield  {author} {\bibinfo {author} {\bibfnamefont {Stefano}\ \bibnamefont
  {Longhi}},\ }\bibfield  {title} {{\selectlanguage {English}\enquote {\bibinfo
  {title} {Parity-time symmetry meets photonics: {{A}} new twist in
  non-{{Hermitian}} optics},}\ }}\href {\doibase 10.1209/0295-5075/120/64001}
  {\bibfield  {journal} {\bibinfo  {journal} {EPL}\ }\textbf {\bibinfo {volume}
  {120}},\ \bibinfo {pages} {64001} (\bibinfo {year} {2017})}\BibitemShut
  {NoStop}%
\bibitem [{\citenamefont {{El-Ganainy}}\ \emph {et~al.}(2018)\citenamefont
  {{El-Ganainy}}, \citenamefont {Makris}, \citenamefont {Khajavikhan},
  \citenamefont {Musslimani}, \citenamefont {Rotter},\ and\ \citenamefont
  {Christodoulides}}]{El-Ganainy:2018}%
  \BibitemOpen
  \bibfield  {author} {\bibinfo {author} {\bibfnamefont {Ramy}\ \bibnamefont
  {{El-Ganainy}}}, \bibinfo {author} {\bibfnamefont {Konstantinos~G.}\
  \bibnamefont {Makris}}, \bibinfo {author} {\bibfnamefont {Mercedeh}\
  \bibnamefont {Khajavikhan}}, \bibinfo {author} {\bibfnamefont {Ziad~H.}\
  \bibnamefont {Musslimani}}, \bibinfo {author} {\bibfnamefont {Stefan}\
  \bibnamefont {Rotter}}, \ and\ \bibinfo {author} {\bibfnamefont
  {Demetrios~N.}\ \bibnamefont {Christodoulides}},\ }\bibfield  {title}
  {{\selectlanguage {English}\enquote {\bibinfo {title} {Non-{{Hermitian}} physics
  and {{PT}} symmetry},}\ }}\href {\doibase 10.1038/nphys4323} {\bibfield
  {journal} {\bibinfo  {journal} {Nat. Phys.}\ }\textbf {\bibinfo {volume}
  {14}},\ \bibinfo {pages} {11--19} (\bibinfo {year} {2018})}\BibitemShut
  {NoStop}%
\bibitem [{\citenamefont {Bahari}\ \emph {et~al.}(2017)\citenamefont {Bahari},
  \citenamefont {Ndao}, \citenamefont {Vallini}, \citenamefont {Amili},
  \citenamefont {Fainman},\ and\ \citenamefont {Kant{\'e}}}]{Bahari:2017}%
  \BibitemOpen
  \bibfield  {author} {\bibinfo {author} {\bibfnamefont {Babak}\ \bibnamefont
  {Bahari}}, \bibinfo {author} {\bibfnamefont {Abdoulaye}\ \bibnamefont
  {Ndao}}, \bibinfo {author} {\bibfnamefont {Felipe}\ \bibnamefont {Vallini}},
  \bibinfo {author} {\bibfnamefont {Abdelkrim~El}\ \bibnamefont {Amili}},
  \bibinfo {author} {\bibfnamefont {Yeshaiahu}\ \bibnamefont {Fainman}}, \ and\
  \bibinfo {author} {\bibfnamefont {Boubacar}\ \bibnamefont {Kant{\'e}}},\
  }\bibfield  {title} {{\selectlanguage {English}\enquote {\bibinfo {title}
  {Nonreciprocal lasing in topological cavities of arbitrary geometries},}\
  }}\href {\doibase 10.1126/science.aao4551} {\bibfield  {journal} {\bibinfo
  {journal} {Science}\ }\textbf {\bibinfo {volume} {358}},\ \bibinfo {pages}
  {636} (\bibinfo {year} {2017})}\BibitemShut {NoStop}%
\bibitem [{\citenamefont {{St-Jean}}\ \emph {et~al.}(2017)\citenamefont
  {{St-Jean}}, \citenamefont {Goblot}, \citenamefont {Galopin}, \citenamefont
  {Lema{\^i}tre}, \citenamefont {Ozawa}, \citenamefont {Gratiet}, \citenamefont
  {Sagnes}, \citenamefont {Bloch},\ and\ \citenamefont {Amo}}]{St-Jean:2017}%
  \BibitemOpen
  \bibfield  {author} {\bibinfo {author} {\bibfnamefont {P.}~\bibnamefont
  {{St-Jean}}}, \bibinfo {author} {\bibfnamefont {V.}~\bibnamefont {Goblot}},
  \bibinfo {author} {\bibfnamefont {E.}~\bibnamefont {Galopin}}, \bibinfo
  {author} {\bibfnamefont {A.}~\bibnamefont {Lema{\^i}tre}}, \bibinfo {author}
  {\bibfnamefont {T.}~\bibnamefont {Ozawa}}, \bibinfo {author} {\bibfnamefont
  {L.~Le}\ \bibnamefont {Gratiet}}, \bibinfo {author} {\bibfnamefont
  {I.}~\bibnamefont {Sagnes}}, \bibinfo {author} {\bibfnamefont
  {J.}~\bibnamefont {Bloch}}, \ and\ \bibinfo {author} {\bibfnamefont
  {A.}~\bibnamefont {Amo}},\ }\bibfield  {title} {{\selectlanguage {English}\enquote
  {\bibinfo {title} {Lasing in topological edge states of a one-dimensional
  lattice},}\ }}\href {\doibase 10.1038/s41566-017-0006-2} {\bibfield
  {journal} {\bibinfo  {journal} {Nat. Photonics}\ }\textbf {\bibinfo {volume}
  {11}},\ \bibinfo {pages} {651} (\bibinfo {year} {2017})}\BibitemShut
  {NoStop}%
\bibitem [{\citenamefont {Parto}\ \emph {et~al.}(2018)\citenamefont {Parto},
  \citenamefont {Wittek}, \citenamefont {Hodaei}, \citenamefont {Harari},
  \citenamefont {Bandres}, \citenamefont {Ren}, \citenamefont {Rechtsman},
  \citenamefont {Segev}, \citenamefont {Christodoulides},\ and\ \citenamefont
  {Khajavikhan}}]{Parto:2018}%
  \BibitemOpen
  \bibfield  {author} {\bibinfo {author} {\bibfnamefont {Midya}\ \bibnamefont
  {Parto}}, \bibinfo {author} {\bibfnamefont {Steffen}\ \bibnamefont {Wittek}},
  \bibinfo {author} {\bibfnamefont {Hossein}\ \bibnamefont {Hodaei}}, \bibinfo
  {author} {\bibfnamefont {Gal}\ \bibnamefont {Harari}}, \bibinfo {author}
  {\bibfnamefont {Miguel~A.}\ \bibnamefont {Bandres}}, \bibinfo {author}
  {\bibfnamefont {Jinhan}\ \bibnamefont {Ren}}, \bibinfo {author}
  {\bibfnamefont {Mikael~C.}\ \bibnamefont {Rechtsman}}, \bibinfo {author}
  {\bibfnamefont {Mordechai}\ \bibnamefont {Segev}}, \bibinfo {author}
  {\bibfnamefont {Demetrios~N.}\ \bibnamefont {Christodoulides}}, \ and\
  \bibinfo {author} {\bibfnamefont {Mercedeh}\ \bibnamefont {Khajavikhan}},\
  }\bibfield  {title} {\enquote {\bibinfo {title} {Edge-{{Mode Lasing}} in {{1D
  Topological Active Arrays}}},}\ }\href {\doibase
  10.1103/PhysRevLett.120.113901} {\bibfield  {journal} {\bibinfo  {journal}
  {Phys. Rev. Lett.}\ }\textbf {\bibinfo {volume} {120}},\ \bibinfo {pages}
  {113901} (\bibinfo {year} {2018})}\BibitemShut {NoStop}%
\bibitem [{\citenamefont {Bergholtz}\ \emph {et~al.}(2019)\citenamefont
  {Bergholtz}, \citenamefont {Budich},\ and\ \citenamefont
  {Kunst}}]{Bergholtz:2019a}%
  \BibitemOpen
  \bibfield  {author} {\bibinfo {author} {\bibfnamefont {Emil~J.}\ \bibnamefont
  {Bergholtz}}, \bibinfo {author} {\bibfnamefont {Jan~Carl}\ \bibnamefont
  {Budich}}, \ and\ \bibinfo {author} {\bibfnamefont {Flore~K.}\ \bibnamefont
  {Kunst}},\ }\bibfield  {title} {\enquote {\bibinfo {title} {Exceptional
  {{Topology}} of {{Non}}-{{Hermitian Systems}}},}\ }\href@noop {} {\bibfield
  {journal} {\bibinfo  {journal} {arXiv:1912.10048}\ } (\bibinfo {year}
  {2019})},\ \Eprint {http://arxiv.org/abs/1912.10048} {arXiv:1912.10048}
  \BibitemShut {NoStop}%
\bibitem [{\citenamefont {Gong}\ \emph {et~al.}(2018)\citenamefont {Gong},
  \citenamefont {Ashida}, \citenamefont {Kawabata}, \citenamefont {Takasan},
  \citenamefont {Higashikawa},\ and\ \citenamefont {Ueda}}]{Gong:2018}%
  \BibitemOpen
  \bibfield  {author} {\bibinfo {author} {\bibfnamefont {Zongping}\
  \bibnamefont {Gong}}, \bibinfo {author} {\bibfnamefont {Yuto}\ \bibnamefont
  {Ashida}}, \bibinfo {author} {\bibfnamefont {Kohei}\ \bibnamefont
  {Kawabata}}, \bibinfo {author} {\bibfnamefont {Kazuaki}\ \bibnamefont
  {Takasan}}, \bibinfo {author} {\bibfnamefont {Sho}\ \bibnamefont
  {Higashikawa}}, \ and\ \bibinfo {author} {\bibfnamefont {Masahito}\
  \bibnamefont {Ueda}},\ }\bibfield  {title} {{\selectlanguage {English}\enquote
  {\bibinfo {title} {Topological phases of non-{{Hermitian}} systems},}\
  }}\href {\doibase 10.1103/PhysRevX.8.031079} {\bibfield  {journal} {\bibinfo
  {journal} {Phys. Rev. X}\ }\textbf {\bibinfo {volume} {8}},\ \bibinfo {pages}
  {031079} (\bibinfo {year} {2018})},\ \Eprint
  {http://arxiv.org/abs/1802.07964} {arXiv:1802.07964} \BibitemShut {NoStop}%
\bibitem [{\citenamefont {Kawabata}\ \emph {et~al.}(2019)\citenamefont
  {Kawabata}, \citenamefont {Shiozaki}, \citenamefont {Ueda},\ and\
  \citenamefont {Sato}}]{Kawabata:2019b}%
  \BibitemOpen
  \bibfield  {author} {\bibinfo {author} {\bibfnamefont {Kohei}\ \bibnamefont
  {Kawabata}}, \bibinfo {author} {\bibfnamefont {Ken}\ \bibnamefont
  {Shiozaki}}, \bibinfo {author} {\bibfnamefont {Masahito}\ \bibnamefont
  {Ueda}}, \ and\ \bibinfo {author} {\bibfnamefont {Masatoshi}\ \bibnamefont
  {Sato}},\ }\bibfield  {title} {\enquote {\bibinfo {title} {Symmetry and
  {{Topology}} in {{Non}}-{{Hermitian Physics}}},}\ }\href {\doibase
  10.1103/PhysRevX.9.041015} {\bibfield  {journal} {\bibinfo  {journal} {Phys.
  Rev. X}\ }\textbf {\bibinfo {volume} {9}},\ \bibinfo {pages} {041015}
  (\bibinfo {year} {2019})}\BibitemShut {NoStop}%
\bibitem [{\citenamefont {Zhou}\ and\ \citenamefont {Lee}(2019)}]{Zhou:2019a}%
  \BibitemOpen
  \bibfield  {author} {\bibinfo {author} {\bibfnamefont {Hengyun}\ \bibnamefont
  {Zhou}}\ and\ \bibinfo {author} {\bibfnamefont {Jong~Yeon}\ \bibnamefont
  {Lee}},\ }\bibfield  {title} {\enquote {\bibinfo {title} {Periodic table for
  topological bands with non-{{Hermitian}} symmetries},}\ }\href {\doibase
  10.1103/PhysRevB.99.235112} {\bibfield  {journal} {\bibinfo  {journal} {Phys.
  Rev. B}\ }\textbf {\bibinfo {volume} {99}},\ \bibinfo {pages} {235112}
  (\bibinfo {year} {2019})},\ \Eprint {http://arxiv.org/abs/1812.10490}
  {arXiv:1812.10490} \BibitemShut {NoStop}%
\bibitem [{\citenamefont {Shnerb}\ and\ \citenamefont
  {Nelson}(1998)}]{Shnerb:1998}%
  \BibitemOpen
  \bibfield  {author} {\bibinfo {author} {\bibfnamefont {Nadav~M.}\
  \bibnamefont {Shnerb}}\ and\ \bibinfo {author} {\bibfnamefont {David~R.}\
  \bibnamefont {Nelson}},\ }\bibfield  {title} {\enquote {\bibinfo {title}
  {Winding {{Numbers}}, {{Complex Currents}}, and {{Non}}-{{Hermitian
  Localization}}},}\ }\href {\doibase 10.1103/PhysRevLett.80.5172} {\bibfield
  {journal} {\bibinfo  {journal} {Phys. Rev. Lett.}\ }\textbf {\bibinfo
  {volume} {80}},\ \bibinfo {pages} {5172--5175} (\bibinfo {year}
  {1998})}\BibitemShut {NoStop}%
\bibitem [{\citenamefont {Lee}(2016)}]{Lee:2016}%
  \BibitemOpen
  \bibfield  {author} {\bibinfo {author} {\bibfnamefont {Tony~E}\ \bibnamefont
  {Lee}},\ }\bibfield  {title} {{\selectlanguage {English}\enquote {\bibinfo
  {title} {Anomalous {{Edge State}} in a {{Non}}-{{Hermitian Lattice}}},}\
  }}\href {\doibase 10.1103/PhysRevLett.116.133903} {\bibfield  {journal}
  {\bibinfo  {journal} {Phys. Rev. Lett.}\ }\textbf {\bibinfo {volume} {116}},\
  \bibinfo {pages} {133903} (\bibinfo {year} {2016})}\BibitemShut {NoStop}%
\bibitem [{\citenamefont {Leykam}\ \emph {et~al.}(2017)\citenamefont {Leykam},
  \citenamefont {Bliokh}, \citenamefont {Huang}, \citenamefont {Chong},\ and\
  \citenamefont {Nori}}]{Leykam:2017}%
  \BibitemOpen
  \bibfield  {author} {\bibinfo {author} {\bibfnamefont {Daniel}\ \bibnamefont
  {Leykam}}, \bibinfo {author} {\bibfnamefont {Konstantin~Y}\ \bibnamefont
  {Bliokh}}, \bibinfo {author} {\bibfnamefont {Chunli}\ \bibnamefont {Huang}},
  \bibinfo {author} {\bibfnamefont {Y~D}\ \bibnamefont {Chong}}, \ and\
  \bibinfo {author} {\bibfnamefont {Franco}\ \bibnamefont {Nori}},\ }\bibfield
  {title} {{\selectlanguage {English}\enquote {\bibinfo {title} {Edge
  {{Modes}}, {{Degeneracies}}, and {{Topological Numbers}} in
  {{Non}}-{{Hermitian Systems}}},}\ }}\href {\doibase
  10.1103/PhysRevLett.118.040401} {\bibfield  {journal} {\bibinfo  {journal}
  {Phys. Rev. Lett.}\ }\textbf {\bibinfo {volume} {118}},\ \bibinfo {pages}
  {040401} (\bibinfo {year} {2017})}\BibitemShut {NoStop}%
\bibitem [{\citenamefont {Shen}\ \emph {et~al.}(2018)\citenamefont {Shen},
  \citenamefont {Zhen},\ and\ \citenamefont {Fu}}]{Shen:2018}%
  \BibitemOpen
  \bibfield  {author} {\bibinfo {author} {\bibfnamefont {Huitao}\ \bibnamefont
  {Shen}}, \bibinfo {author} {\bibfnamefont {Bo}~\bibnamefont {Zhen}}, \ and\
  \bibinfo {author} {\bibfnamefont {Liang}\ \bibnamefont {Fu}},\ }\bibfield
  {title} {{\selectlanguage {English}\enquote {\bibinfo {title} {Topological
  {{Band Theory}} for {{Non}}-{{Hermitian Hamiltonians}}},}\ }}\href {\doibase
  10.1103/PhysRevA.72.014104} {\bibfield  {journal} {\bibinfo  {journal} {Phys.
  Rev. Lett.}\ }\textbf {\bibinfo {volume} {120}},\ \bibinfo {pages} {146402}
  (\bibinfo {year} {2018})}\BibitemShut {NoStop}%
\bibitem [{\citenamefont {Lieu}(2018{\natexlab{a}})}]{Lieu:2018}%
  \BibitemOpen
  \bibfield  {author} {\bibinfo {author} {\bibfnamefont {Simon}\ \bibnamefont
  {Lieu}},\ }\bibfield  {title} {\enquote {\bibinfo {title} {Topological
  symmetry classes for non-{{Hermitian}} models and connections to the bosonic
  {{Bogoliubov}}--de {{Gennes}} equation},}\ }\href {\doibase
  10.1103/PhysRevB.98.115135} {\bibfield  {journal} {\bibinfo  {journal} {Phys.
  Rev. B}\ }\textbf {\bibinfo {volume} {98}},\ \bibinfo {pages} {115135}
  (\bibinfo {year} {2018}{\natexlab{a}})}\BibitemShut {NoStop}%
\bibitem [{\citenamefont {Hirsbrunner}\ \emph {et~al.}(2019)\citenamefont
  {Hirsbrunner}, \citenamefont {Philip},\ and\ \citenamefont
  {Gilbert}}]{Hirsbrunner:2019}%
  \BibitemOpen
  \bibfield  {author} {\bibinfo {author} {\bibfnamefont {Mark~R.}\ \bibnamefont
  {Hirsbrunner}}, \bibinfo {author} {\bibfnamefont {Timothy~M.}\ \bibnamefont
  {Philip}}, \ and\ \bibinfo {author} {\bibfnamefont {Matthew~J.}\ \bibnamefont
  {Gilbert}},\ }\bibfield  {title} {\enquote {\bibinfo {title} {Topology and
  observables of the non-{{Hermitian Chern}} insulator},}\ }\href {\doibase
  10.1103/PhysRevB.100.081104} {\bibfield  {journal} {\bibinfo  {journal}
  {Phys. Rev. B}\ }\textbf {\bibinfo {volume} {100}},\ \bibinfo {pages}
  {081104(R)} (\bibinfo {year} {2019})},\ \Eprint
  {http://arxiv.org/abs/1901.09961} {arXiv:1901.09961} \BibitemShut {NoStop}%
\bibitem [{\citenamefont {Longhi}(2019{\natexlab{a}})}]{Longhi:2019a}%
  \BibitemOpen
  \bibfield  {author} {\bibinfo {author} {\bibfnamefont {S.}~\bibnamefont
  {Longhi}},\ }\bibfield  {title} {\enquote {\bibinfo {title} {Topological
  {{Phase Transition}} in non-{{Hermitian Quasicrystals}}},}\ }\href {\doibase
  10.1103/PhysRevLett.122.237601} {\bibfield  {journal} {\bibinfo  {journal}
  {Phys. Rev. Lett.}\ }\textbf {\bibinfo {volume} {122}},\ \bibinfo {pages}
  {237601} (\bibinfo {year} {2019}{\natexlab{a}})},\ \Eprint
  {http://arxiv.org/abs/1905.09460} {arXiv:1905.09460} \BibitemShut {NoStop}%
\bibitem [{\citenamefont {Chen}\ and\ \citenamefont {Zhai}(2018)}]{Chen:2018c}%
  \BibitemOpen
  \bibfield  {author} {\bibinfo {author} {\bibfnamefont {Yu}~\bibnamefont
  {Chen}}\ and\ \bibinfo {author} {\bibfnamefont {Hui}\ \bibnamefont {Zhai}},\
  }\bibfield  {title} {\enquote {\bibinfo {title} {Hall conductance of a
  non-{{Hermitian Chern}} insulator},}\ }\href {\doibase
  10.1103/PhysRevB.98.245130} {\bibfield  {journal} {\bibinfo  {journal} {Phys.
  Rev. B}\ }\textbf {\bibinfo {volume} {98}},\ \bibinfo {pages} {245130}
  (\bibinfo {year} {2018})}\BibitemShut {NoStop}%
\bibitem [{\citenamefont {Xiong}(2018)}]{Xiong:2018}%
  \BibitemOpen
  \bibfield  {author} {\bibinfo {author} {\bibfnamefont {Ye}~\bibnamefont
  {Xiong}},\ }\bibfield  {title} {\enquote {\bibinfo {title} {Why does bulk
  boundary correspondence fail in some non-hermitian topological models},}\
  }\href {\doibase 10.1088/2399-6528/aab64a} {\bibfield  {journal} {\bibinfo
  {journal} {J. Phys. Commun.}\ }\textbf {\bibinfo {volume} {2}},\ \bibinfo
  {pages} {035043} (\bibinfo {year} {2018})}\BibitemShut {NoStop}%
\bibitem [{\citenamefont {Kunst}\ \emph {et~al.}(2018)\citenamefont {Kunst},
  \citenamefont {Edvardsson}, \citenamefont {Budich},\ and\ \citenamefont
  {Bergholtz}}]{Kunst:2018}%
  \BibitemOpen
  \bibfield  {author} {\bibinfo {author} {\bibfnamefont {Flore~K}\ \bibnamefont
  {Kunst}}, \bibinfo {author} {\bibfnamefont {Elisabet}\ \bibnamefont
  {Edvardsson}}, \bibinfo {author} {\bibfnamefont {Jan~Carl}\ \bibnamefont
  {Budich}}, \ and\ \bibinfo {author} {\bibfnamefont {Emil~J}\ \bibnamefont
  {Bergholtz}},\ }\bibfield  {title} {{\selectlanguage {English}\enquote
  {\bibinfo {title} {Biorthogonal {{Bulk}}-{{Boundary Correspondence}} in
  {{Non}}-{{Hermitian Systems}}},}\ }}\href {\doibase
  10.1103/PhysRevLett.121.026808} {\bibfield  {journal} {\bibinfo  {journal}
  {Phys. Rev. Lett.}\ }\textbf {\bibinfo {volume} {121}},\ \bibinfo {pages}
  {026808} (\bibinfo {year} {2018})}\BibitemShut {NoStop}%
\bibitem [{\citenamefont {Yao}\ and\ \citenamefont {Wang}(2018)}]{Yao:2018b}%
  \BibitemOpen
  \bibfield  {author} {\bibinfo {author} {\bibfnamefont {Shunyu}\ \bibnamefont
  {Yao}}\ and\ \bibinfo {author} {\bibfnamefont {Zhong}\ \bibnamefont {Wang}},\
  }\bibfield  {title} {{\selectlanguage {English}\enquote {\bibinfo {title} {Edge
  {{States}} and {{Topological Invariants}} of {{Non}}-{{Hermitian
  Systems}}},}\ }}\href {\doibase 10.1103/PhysRevLett.121.086803} {\bibfield
  {journal} {\bibinfo  {journal} {Phys. Rev. Lett.}\ }\textbf {\bibinfo
  {volume} {121}},\ \bibinfo {pages} {086803} (\bibinfo {year}
  {2018})}\BibitemShut {NoStop}%
\bibitem [{\citenamefont {Lee}\ and\ \citenamefont
  {Thomale}(2019)}]{Lee:2019c}%
  \BibitemOpen
  \bibfield  {author} {\bibinfo {author} {\bibfnamefont {Ching~Hua}\
  \bibnamefont {Lee}}\ and\ \bibinfo {author} {\bibfnamefont {Ronny}\
  \bibnamefont {Thomale}},\ }\bibfield  {title} {\enquote {\bibinfo {title}
  {Anatomy of skin modes and topology in non-{{Hermitian}} systems},}\ }\href
  {\doibase 10.1103/PhysRevB.99.201103} {\bibfield  {journal} {\bibinfo
  {journal} {Phys. Rev. B}\ }\textbf {\bibinfo {volume} {99}},\ \bibinfo
  {pages} {201103(R)} (\bibinfo {year} {2019})}\BibitemShut {NoStop}%
\bibitem [{\citenamefont {Jin}\ and\ \citenamefont {Song}(2019)}]{Jin:2019a}%
  \BibitemOpen
  \bibfield  {author} {\bibinfo {author} {\bibfnamefont {L.}~\bibnamefont
  {Jin}}\ and\ \bibinfo {author} {\bibfnamefont {Z.}~\bibnamefont {Song}},\
  }\bibfield  {title} {\enquote {\bibinfo {title} {Bulk-boundary correspondence
  in a non-{{Hermitian}} system in one dimension with chiral inversion
  symmetry},}\ }\href {\doibase 10.1103/PhysRevB.99.081103} {\bibfield
  {journal} {\bibinfo  {journal} {Phys. Rev. B}\ }\textbf {\bibinfo {volume}
  {99}},\ \bibinfo {pages} {081103(R)} (\bibinfo {year} {2019})},\ \Eprint
  {http://arxiv.org/abs/1809.03139} {arXiv:1809.03139} \BibitemShut {NoStop}%
\bibitem [{\citenamefont {Herviou}\ \emph {et~al.}(2019)\citenamefont
  {Herviou}, \citenamefont {Bardarson},\ and\ \citenamefont
  {Regnault}}]{Herviou:2019}%
  \BibitemOpen
  \bibfield  {author} {\bibinfo {author} {\bibfnamefont {Lo{\"i}c}\
  \bibnamefont {Herviou}}, \bibinfo {author} {\bibfnamefont {Jens~H.}\
  \bibnamefont {Bardarson}}, \ and\ \bibinfo {author} {\bibfnamefont {Nicolas}\
  \bibnamefont {Regnault}},\ }\bibfield  {title} {\enquote {\bibinfo {title}
  {Defining a bulk-edge correspondence for non-{{Hermitian Hamiltonians}} via
  singular-value decomposition},}\ }\href {\doibase 10.1103/PhysRevA.99.052118}
  {\bibfield  {journal} {\bibinfo  {journal} {Phys. Rev. A}\ }\textbf {\bibinfo
  {volume} {99}},\ \bibinfo {pages} {052118} (\bibinfo {year}
  {2019})}\BibitemShut {NoStop}%
\bibitem [{\citenamefont {Ge}\ \emph {et~al.}(2019)\citenamefont {Ge},
  \citenamefont {Zhang}, \citenamefont {Liu}, \citenamefont {Li}, \citenamefont
  {Fan},\ and\ \citenamefont {Nori}}]{Ge:2019}%
  \BibitemOpen
  \bibfield  {author} {\bibinfo {author} {\bibfnamefont {Zi-Yong}\ \bibnamefont
  {Ge}}, \bibinfo {author} {\bibfnamefont {Yu-Ran}\ \bibnamefont {Zhang}},
  \bibinfo {author} {\bibfnamefont {Tao}\ \bibnamefont {Liu}}, \bibinfo
  {author} {\bibfnamefont {Si-Wen}\ \bibnamefont {Li}}, \bibinfo {author}
  {\bibfnamefont {Heng}\ \bibnamefont {Fan}}, \ and\ \bibinfo {author}
  {\bibfnamefont {Franco}\ \bibnamefont {Nori}},\ }\bibfield  {title} {\enquote
  {\bibinfo {title} {Topological band theory for non-{{Hermitian}} systems from
  the {{Dirac}} equation},}\ }\href {\doibase 10.1103/PhysRevB.100.054105}
  {\bibfield  {journal} {\bibinfo  {journal} {Phys. Rev. B}\ }\textbf {\bibinfo
  {volume} {100}},\ \bibinfo {pages} {054105} (\bibinfo {year}
  {2019})}\BibitemShut {NoStop}%
\bibitem [{\citenamefont {Borgnia}\ \emph {et~al.}(2020)\citenamefont
  {Borgnia}, \citenamefont {Kruchkov},\ and\ \citenamefont
  {Slager}}]{Borgnia:2020}%
  \BibitemOpen
  \bibfield  {author} {\bibinfo {author} {\bibfnamefont {Dan~S.}\ \bibnamefont
  {Borgnia}}, \bibinfo {author} {\bibfnamefont {Alex~Jura}\ \bibnamefont
  {Kruchkov}}, \ and\ \bibinfo {author} {\bibfnamefont {Robert-Jan}\
  \bibnamefont {Slager}},\ }\bibfield  {title} {\enquote {\bibinfo {title}
  {Non-{{Hermitian Boundary Modes}} and {{Topology}}},}\ }\href {\doibase
  10.1103/PhysRevLett.124.056802} {\bibfield  {journal} {\bibinfo  {journal}
  {Phys. Rev. Lett.}\ }\textbf {\bibinfo {volume} {124}},\ \bibinfo {pages}
  {056802} (\bibinfo {year} {2020})},\ \Eprint
  {http://arxiv.org/abs/1902.07217} {arXiv:1902.07217} \BibitemShut {NoStop}%
\bibitem [{\citenamefont {Brzezicki}\ and\ \citenamefont
  {Hyart}(2019)}]{Brzezicki:2019a}%
  \BibitemOpen
  \bibfield  {author} {\bibinfo {author} {\bibfnamefont {Wojciech}\
  \bibnamefont {Brzezicki}}\ and\ \bibinfo {author} {\bibfnamefont {Timo}\
  \bibnamefont {Hyart}},\ }\bibfield  {title} {\enquote {\bibinfo {title}
  {Hidden {{Chern}} number in one-dimensional non-{{Hermitian}}
  chiral-symmetric systems},}\ }\href {\doibase 10.1103/PhysRevB.100.161105}
  {\bibfield  {journal} {\bibinfo  {journal} {Phys. Rev. B}\ }\textbf {\bibinfo
  {volume} {100}},\ \bibinfo {pages} {161105(R)} (\bibinfo {year}
  {2019})}\BibitemShut {NoStop}%
\bibitem [{\citenamefont {Yokomizo}\ and\ \citenamefont
  {Murakami}(2019)}]{Yokomizo:2019}%
  \BibitemOpen
  \bibfield  {author} {\bibinfo {author} {\bibfnamefont {Kazuki}\ \bibnamefont
  {Yokomizo}}\ and\ \bibinfo {author} {\bibfnamefont {Shuichi}\ \bibnamefont
  {Murakami}},\ }\bibfield  {title} {\enquote {\bibinfo {title} {Non-{{Bloch
  Band Theory}} of {{Non}}-{{Hermitian Systems}}},}\ }\href {\doibase
  10.1103/PhysRevLett.123.066404} {\bibfield  {journal} {\bibinfo  {journal}
  {Phys. Rev. Lett.}\ }\textbf {\bibinfo {volume} {123}},\ \bibinfo {pages}
  {066404} (\bibinfo {year} {2019})}\BibitemShut {NoStop}%
\bibitem [{\citenamefont {Kawabata}\ \emph {et~al.}(2020)\citenamefont
  {Kawabata}, \citenamefont {Okuma},\ and\ \citenamefont
  {Sato}}]{Kawabata:2020}%
  \BibitemOpen
  \bibfield  {author} {\bibinfo {author} {\bibfnamefont {Kohei}\ \bibnamefont
  {Kawabata}}, \bibinfo {author} {\bibfnamefont {Nobuyuki}\ \bibnamefont
  {Okuma}}, \ and\ \bibinfo {author} {\bibfnamefont {Masatoshi}\ \bibnamefont
  {Sato}},\ }\bibfield  {title} {\enquote {\bibinfo {title} {Non-{{Bloch}} band
  theory of non-{{Hermitian Hamiltonians}} in the symplectic class},}\ }\href
  {\doibase 10.1103/PhysRevB.101.195147} {\bibfield  {journal} {\bibinfo
  {journal} {Phys. Rev. B}\ }\textbf {\bibinfo {volume} {101}},\ \bibinfo
  {pages} {195147} (\bibinfo {year} {2020})},\ \Eprint
  {http://arxiv.org/abs/2003.07597} {arXiv:2003.07597} \BibitemShut {NoStop}%
\bibitem [{\citenamefont {Yang}\ \emph {et~al.}(2019)\citenamefont {Yang},
  \citenamefont {Zhang}, \citenamefont {Fang},\ and\ \citenamefont
  {Hu}}]{Yang:2019g}%
  \BibitemOpen
  \bibfield  {author} {\bibinfo {author} {\bibfnamefont {Zhesen}\ \bibnamefont
  {Yang}}, \bibinfo {author} {\bibfnamefont {Kai}\ \bibnamefont {Zhang}},
  \bibinfo {author} {\bibfnamefont {Chen}\ \bibnamefont {Fang}}, \ and\
  \bibinfo {author} {\bibfnamefont {Jiangping}\ \bibnamefont {Hu}},\ }\bibfield
   {title} {\enquote {\bibinfo {title} {Auxiliary generalized {{Brillouin}}
  zone method in non-{{Hermitian}} band theory},}\ }\href@noop {} {\bibfield
  {journal} {\bibinfo  {journal} {arXiv:1912.05499}\ } (\bibinfo {year}
  {2019})},\ \Eprint {http://arxiv.org/abs/1912.05499} {arXiv:1912.05499}
  \BibitemShut {NoStop}%
\bibitem [{Note29()}]{Note29}%
  \BibitemOpen
  \bibinfo {note} {See Supplemental Material at \protect \dots {} for
  non-Hermitian extensions of the SSH lattice model and an elementary
  discussion of the Mahaux-Weidenm{\"u}ller formula.}\BibitemShut {Stop}%
\bibitem [{\citenamefont {Schnyder}\ \emph {et~al.}(2008)\citenamefont
  {Schnyder}, \citenamefont {Ryu}, \citenamefont {Furusaki},\ and\
  \citenamefont {Ludwig}}]{Schnyder:2008}%
  \BibitemOpen
  \bibfield  {author} {\bibinfo {author} {\bibfnamefont {Andreas~P}\
  \bibnamefont {Schnyder}}, \bibinfo {author} {\bibfnamefont {Shinsei}\
  \bibnamefont {Ryu}}, \bibinfo {author} {\bibfnamefont {Akira}\ \bibnamefont
  {Furusaki}}, \ and\ \bibinfo {author} {\bibfnamefont {Andreas W~W}\
  \bibnamefont {Ludwig}},\ }\bibfield  {title} {\enquote {\bibinfo {title}
  {Classification of topological insulators and superconductors in three
  spatial dimensions},}\ }\href {\doibase 10.1103/PhysRevB.78.195125}
  {\bibfield  {journal} {\bibinfo  {journal} {Phys. Rev. B}\ }\textbf {\bibinfo
  {volume} {78}},\ \bibinfo {pages} {195125} (\bibinfo {year}
  {2008})}\BibitemShut {NoStop}%
\bibitem [{\citenamefont {Kitaev}(2009)}]{Kitaev:2009}%
  \BibitemOpen
  \bibfield  {author} {\bibinfo {author} {\bibfnamefont {Alexei}\ \bibnamefont
  {Kitaev}},\ }\bibfield  {title} {\enquote {\bibinfo {title} {Periodic table
  for topological insulators and superconductors},}\ }in\ \href {\doibase
  10.1063/1.3149495} {\emph {\bibinfo {booktitle} {{{AIP Conf}}. {{Proc}}.}}}\
  (\bibinfo {year} {2009})\ p.~\bibinfo {pages} {22}\BibitemShut {NoStop}%
\bibitem [{\citenamefont {Zirnstein}\ and\ \citenamefont
  {Rosenow}(2020)}]{Zirnstein:2020}%
  \BibitemOpen
  \bibfield  {author} {\bibinfo {author} {\bibfnamefont {Heinrich-Gregor}\
  \bibnamefont {Zirnstein}}\ and\ \bibinfo {author} {\bibfnamefont {Bernd}\
  \bibnamefont {Rosenow}},\ }\bibfield  {title} {\enquote {\bibinfo {title}
  {Exponentially growing bulk {{Green}} functions as signature of nontrivial
  non-{{Hermitian}} winding number in one dimension},}\ }\href@noop {}
  {\bibfield  {journal} {\bibinfo  {journal} {arXiv:2007.07026}\ } (\bibinfo
  {year} {2020})},\ \Eprint {http://arxiv.org/abs/2007.07026}
  {arXiv:2007.07026} \BibitemShut {NoStop}%
\bibitem [{\citenamefont {Martinez~Alvarez}\ \emph
  {et~al.}(2018{\natexlab{a}})\citenamefont {Martinez~Alvarez}, \citenamefont
  {Barrios~Vargas},\ and\ \citenamefont {Foa~Torres}}]{MartinezAlvarez:2018}%
  \BibitemOpen
  \bibfield  {author} {\bibinfo {author} {\bibfnamefont {V.~M.}\ \bibnamefont
  {Martinez~Alvarez}}, \bibinfo {author} {\bibfnamefont {J.~E.}\ \bibnamefont
  {Barrios~Vargas}}, \ and\ \bibinfo {author} {\bibfnamefont {L.~E.~F.}\
  \bibnamefont {Foa~Torres}},\ }\bibfield  {title} {{\selectlanguage
  {English}\enquote {\bibinfo {title} {Non-{{Hermitian}} robust edge states in one
  dimension: {{Anomalous}} localization and eigenspace condensation at
  exceptional points},}\ }}\href {\doibase 10.1103/PhysRevB.97.121401}
  {\bibfield  {journal} {\bibinfo  {journal} {Phys. Rev. B}\ }\textbf {\bibinfo
  {volume} {97}},\ \bibinfo {pages} {121401(R)} (\bibinfo {year}
  {2018}{\natexlab{a}})}\BibitemShut {NoStop}%
\bibitem [{\citenamefont {Martinez~Alvarez}\ \emph
  {et~al.}(2018{\natexlab{b}})\citenamefont {Martinez~Alvarez}, \citenamefont
  {Barrios~Vargas}, \citenamefont {Berdakin},\ and\ \citenamefont
  {Foa~Torres}}]{MartinezAlvarez:2018a}%
  \BibitemOpen
  \bibfield  {author} {\bibinfo {author} {\bibfnamefont {V.~M.}\ \bibnamefont
  {Martinez~Alvarez}}, \bibinfo {author} {\bibfnamefont {J.~E.}\ \bibnamefont
  {Barrios~Vargas}}, \bibinfo {author} {\bibfnamefont {M.}~\bibnamefont
  {Berdakin}}, \ and\ \bibinfo {author} {\bibfnamefont {L.~E.~F.}\ \bibnamefont
  {Foa~Torres}},\ }\bibfield  {title} {{\selectlanguage {English}\enquote {\bibinfo
  {title} {Topological states of non-{{Hermitian}} systems},}\ }}\href
  {\doibase 10.1140/epjst/e2018-800091-5} {\bibfield  {journal} {\bibinfo
  {journal} {Eur. Phys. J. Spec. Top.}\ } (\bibinfo {year}
  {2018}{\natexlab{b}}),\ 10.1140/epjst/e2018-800091-5}\BibitemShut {NoStop}%
\bibitem [{\citenamefont {Zhang}\ \emph {et~al.}(2020)\citenamefont {Zhang},
  \citenamefont {Yang},\ and\ \citenamefont {Fang}}]{Zhang:2020e}%
  \BibitemOpen
  \bibfield  {author} {\bibinfo {author} {\bibfnamefont {Kai}\ \bibnamefont
  {Zhang}}, \bibinfo {author} {\bibfnamefont {Zhesen}\ \bibnamefont {Yang}}, \
  and\ \bibinfo {author} {\bibfnamefont {Chen}\ \bibnamefont {Fang}},\
  }\bibfield  {title} {\enquote {\bibinfo {title} {Correspondence between
  winding numbers and skin modes in non-hermitian systems},}\ }\href@noop {}
  {\bibfield  {journal} {\bibinfo  {journal} {arXiv:1910.01131}\ } (\bibinfo
  {year} {2020})},\ \Eprint {http://arxiv.org/abs/1910.01131}
  {arXiv:1910.01131} \BibitemShut {NoStop}%
\bibitem [{\citenamefont {Okuma}\ \emph {et~al.}(2020)\citenamefont {Okuma},
  \citenamefont {Kawabata}, \citenamefont {Shiozaki},\ and\ \citenamefont
  {Sato}}]{Okuma:2020}%
  \BibitemOpen
  \bibfield  {author} {\bibinfo {author} {\bibfnamefont {Nobuyuki}\
  \bibnamefont {Okuma}}, \bibinfo {author} {\bibfnamefont {Kohei}\ \bibnamefont
  {Kawabata}}, \bibinfo {author} {\bibfnamefont {Ken}\ \bibnamefont
  {Shiozaki}}, \ and\ \bibinfo {author} {\bibfnamefont {Masatoshi}\
  \bibnamefont {Sato}},\ }\bibfield  {title} {\enquote {\bibinfo {title}
  {Topological {{Origin}} of {{Non}}-{{Hermitian Skin Effects}}},}\ }\href
  {\doibase 10.1103/PhysRevLett.124.086801} {\bibfield  {journal} {\bibinfo
  {journal} {Phys. Rev. Lett.}\ }\textbf {\bibinfo {volume} {124}},\ \bibinfo
  {pages} {086801} (\bibinfo {year} {2020})},\ \Eprint
  {http://arxiv.org/abs/1910.02878} {arXiv:1910.02878} \BibitemShut {NoStop}%
\bibitem [{\citenamefont {Okuma}\ and\ \citenamefont
  {Sato}(2020)}]{Okuma:2020a}%
  \BibitemOpen
  \bibfield  {author} {\bibinfo {author} {\bibfnamefont {Nobuyuki}\
  \bibnamefont {Okuma}}\ and\ \bibinfo {author} {\bibfnamefont {Masatoshi}\
  \bibnamefont {Sato}},\ }\bibfield  {title} {\enquote {\bibinfo {title}
  {Hermitian zero modes protected by nonnormality: {{Application}} of
  pseudospectra},}\ }\href {\doibase 10.1103/PhysRevB.102.014203} {\bibfield
  {journal} {\bibinfo  {journal} {Phys. Rev. B}\ }\textbf {\bibinfo {volume}
  {102}},\ \bibinfo {pages} {014203} (\bibinfo {year} {2020})},\ \Eprint
  {http://arxiv.org/abs/2005.01704} {arXiv:2005.01704} \BibitemShut {NoStop}%
\bibitem [{\citenamefont {Xiao}\ \emph {et~al.}(2020)\citenamefont {Xiao},
  \citenamefont {Deng}, \citenamefont {Wang}, \citenamefont {Zhu},
  \citenamefont {Wang}, \citenamefont {Yi},\ and\ \citenamefont
  {Xue}}]{Xiao:2020}%
  \BibitemOpen
  \bibfield  {author} {\bibinfo {author} {\bibfnamefont {Lei}\ \bibnamefont
  {Xiao}}, \bibinfo {author} {\bibfnamefont {Tianshu}\ \bibnamefont {Deng}},
  \bibinfo {author} {\bibfnamefont {Kunkun}\ \bibnamefont {Wang}}, \bibinfo
  {author} {\bibfnamefont {Gaoyan}\ \bibnamefont {Zhu}}, \bibinfo {author}
  {\bibfnamefont {Zhong}\ \bibnamefont {Wang}}, \bibinfo {author}
  {\bibfnamefont {Wei}\ \bibnamefont {Yi}}, \ and\ \bibinfo {author}
  {\bibfnamefont {Peng}\ \bibnamefont {Xue}},\ }\bibfield  {title}
  {{\selectlanguage {English}\enquote {\bibinfo {title} {Non-{{Hermitian}}
  bulk\textendash boundary correspondence in quantum dynamics},}\ }}\href
  {\doibase 10.1038/s41567-020-0836-6} {\bibfield  {journal} {\bibinfo
  {journal} {Nat. Phys.}\ } (\bibinfo {year} {2020}),\
  10.1038/s41567-020-0836-6},\ \Eprint {http://arxiv.org/abs/1907.12566}
  {arXiv:1907.12566} \BibitemShut {NoStop}%
\bibitem [{\citenamefont {Helbig}\ \emph {et~al.}(2020)\citenamefont {Helbig},
  \citenamefont {Hofmann}, \citenamefont {Imhof}, \citenamefont {Abdelghany},
  \citenamefont {Kiessling}, \citenamefont {Molenkamp}, \citenamefont {Lee},
  \citenamefont {Szameit}, \citenamefont {Greiter},\ and\ \citenamefont
  {Thomale}}]{Helbig:2020}%
  \BibitemOpen
  \bibfield  {author} {\bibinfo {author} {\bibfnamefont {T.}~\bibnamefont
  {Helbig}}, \bibinfo {author} {\bibfnamefont {T.}~\bibnamefont {Hofmann}},
  \bibinfo {author} {\bibfnamefont {S.}~\bibnamefont {Imhof}}, \bibinfo
  {author} {\bibfnamefont {M.}~\bibnamefont {Abdelghany}}, \bibinfo {author}
  {\bibfnamefont {T.}~\bibnamefont {Kiessling}}, \bibinfo {author}
  {\bibfnamefont {L.~W.}\ \bibnamefont {Molenkamp}}, \bibinfo {author}
  {\bibfnamefont {C.~H.}\ \bibnamefont {Lee}}, \bibinfo {author} {\bibfnamefont
  {A.}~\bibnamefont {Szameit}}, \bibinfo {author} {\bibfnamefont
  {M.}~\bibnamefont {Greiter}}, \ and\ \bibinfo {author} {\bibfnamefont
  {R.}~\bibnamefont {Thomale}},\ }\bibfield  {title} {{\selectlanguage
  {English}\enquote {\bibinfo {title} {Generalized bulk\textendash boundary
  correspondence in non-{{Hermitian}} topolectrical circuits},}\ }}\href
  {\doibase 10.1038/s41567-020-0922-9} {\bibfield  {journal} {\bibinfo
  {journal} {Nat. Phys.}\ ,\ \bibinfo {pages} {1--4}} (\bibinfo {year}
  {2020})},\ \Eprint {http://arxiv.org/abs/1907.11562} {arXiv:1907.11562}
  \BibitemShut {NoStop}%
\bibitem [{\citenamefont {Hofmann}\ \emph {et~al.}(2020)\citenamefont
  {Hofmann}, \citenamefont {Helbig}, \citenamefont {Schindler}, \citenamefont
  {Salgo}, \citenamefont {Brzezi{\'n}ska}, \citenamefont {Greiter},
  \citenamefont {Kiessling}, \citenamefont {Wolf}, \citenamefont {Vollhardt},
  \citenamefont {Kaba{\v s}i}, \citenamefont {Lee}, \citenamefont {Bilu{\v
  s}i{\'c}}, \citenamefont {Thomale},\ and\ \citenamefont
  {Neupert}}]{Hofmann:2020}%
  \BibitemOpen
  \bibfield  {author} {\bibinfo {author} {\bibfnamefont {Tobias}\ \bibnamefont
  {Hofmann}}, \bibinfo {author} {\bibfnamefont {Tobias}\ \bibnamefont
  {Helbig}}, \bibinfo {author} {\bibfnamefont {Frank}\ \bibnamefont
  {Schindler}}, \bibinfo {author} {\bibfnamefont {Nora}\ \bibnamefont {Salgo}},
  \bibinfo {author} {\bibfnamefont {Marta}\ \bibnamefont {Brzezi{\'n}ska}},
  \bibinfo {author} {\bibfnamefont {Martin}\ \bibnamefont {Greiter}}, \bibinfo
  {author} {\bibfnamefont {Tobias}\ \bibnamefont {Kiessling}}, \bibinfo
  {author} {\bibfnamefont {David}\ \bibnamefont {Wolf}}, \bibinfo {author}
  {\bibfnamefont {Achim}\ \bibnamefont {Vollhardt}}, \bibinfo {author}
  {\bibfnamefont {Anton}\ \bibnamefont {Kaba{\v s}i}}, \bibinfo {author}
  {\bibfnamefont {Ching~Hua}\ \bibnamefont {Lee}}, \bibinfo {author}
  {\bibfnamefont {Ante}\ \bibnamefont {Bilu{\v s}i{\'c}}}, \bibinfo {author}
  {\bibfnamefont {Ronny}\ \bibnamefont {Thomale}}, \ and\ \bibinfo {author}
  {\bibfnamefont {Titus}\ \bibnamefont {Neupert}},\ }\bibfield  {title}
  {\enquote {\bibinfo {title} {Reciprocal skin effect and its realization in a
  topolectrical circuit},}\ }\href {\doibase 10.1103/PhysRevResearch.2.023265}
  {\bibfield  {journal} {\bibinfo  {journal} {Phys. Rev. Research}\ }\textbf
  {\bibinfo {volume} {2}},\ \bibinfo {pages} {023265} (\bibinfo {year}
  {2020})},\ \Eprint {http://arxiv.org/abs/1908.02759} {arXiv:1908.02759}
  \BibitemShut {NoStop}%
\bibitem [{\citenamefont {Yoshida}\ \emph {et~al.}(2020)\citenamefont
  {Yoshida}, \citenamefont {Mizoguchi},\ and\ \citenamefont
  {Hatsugai}}]{Yoshida:2020}%
  \BibitemOpen
  \bibfield  {author} {\bibinfo {author} {\bibfnamefont {Tsuneya}\ \bibnamefont
  {Yoshida}}, \bibinfo {author} {\bibfnamefont {Tomonari}\ \bibnamefont
  {Mizoguchi}}, \ and\ \bibinfo {author} {\bibfnamefont {Yasuhiro}\
  \bibnamefont {Hatsugai}},\ }\bibfield  {title} {\enquote {\bibinfo {title}
  {Mirror skin effect and its electric circuit simulation},}\ }\href {\doibase
  10.1103/PhysRevResearch.2.022062} {\bibfield  {journal} {\bibinfo  {journal}
  {Phys. Rev. Research}\ }\textbf {\bibinfo {volume} {2}},\ \bibinfo {pages}
  {022062} (\bibinfo {year} {2020})},\ \Eprint
  {http://arxiv.org/abs/1912.12022} {arXiv:1912.12022} \BibitemShut {NoStop}%
\bibitem [{\citenamefont {Longhi}(2019{\natexlab{b}})}]{Longhi:2019d}%
  \BibitemOpen
  \bibfield  {author} {\bibinfo {author} {\bibfnamefont {Stefano}\ \bibnamefont
  {Longhi}},\ }\bibfield  {title} {\enquote {\bibinfo {title} {Probing
  non-{{Hermitian}} skin effect and non-{{Bloch}} phase transitions},}\ }\href
  {\doibase 10.1103/PhysRevResearch.1.023013} {\bibfield  {journal} {\bibinfo
  {journal} {Phys. Rev. Research}\ }\textbf {\bibinfo {volume} {1}},\ \bibinfo
  {pages} {023013} (\bibinfo {year} {2019}{\natexlab{b}})}\BibitemShut
  {NoStop}%
\bibitem [{\citenamefont {Lieu}(2018{\natexlab{b}})}]{Lieu:2018a}%
  \BibitemOpen
  \bibfield  {author} {\bibinfo {author} {\bibfnamefont {Simon}\ \bibnamefont
  {Lieu}},\ }\bibfield  {title} {{\selectlanguage {English}\enquote {\bibinfo
  {title} {Topological phases in the non-{{Hermitian
  Su}}-{{Schrieffer}}-{{Heeger}} model},}\ }}\href {\doibase
  10.1103/PhysRevB.70.125429} {\bibfield  {journal} {\bibinfo  {journal} {Phys.
  Rev. B}\ }\textbf {\bibinfo {volume} {97}},\ \bibinfo {pages} {045106}
  (\bibinfo {year} {2018}{\natexlab{b}})}\BibitemShut {NoStop}%
\bibitem [{\citenamefont {Esaki}\ \emph {et~al.}(2011)\citenamefont {Esaki},
  \citenamefont {Sato}, \citenamefont {Hasebe},\ and\ \citenamefont
  {Kohmoto}}]{Esaki:2011}%
  \BibitemOpen
  \bibfield  {author} {\bibinfo {author} {\bibfnamefont {Kenta}\ \bibnamefont
  {Esaki}}, \bibinfo {author} {\bibfnamefont {Masatoshi}\ \bibnamefont {Sato}},
  \bibinfo {author} {\bibfnamefont {Kazuki}\ \bibnamefont {Hasebe}}, \ and\
  \bibinfo {author} {\bibfnamefont {Mahito}\ \bibnamefont {Kohmoto}},\
  }\bibfield  {title} {{\selectlanguage {English}\enquote {\bibinfo {title}
  {Edge states and topological phases in non-{{Hermitian}} systems},}\ }}\href
  {\doibase 10.1103/PhysRevB.84.205128} {\bibfield  {journal} {\bibinfo
  {journal} {Phys. Rev. B}\ }\textbf {\bibinfo {volume} {84}},\ \bibinfo
  {pages} {205128} (\bibinfo {year} {2011})}\BibitemShut {NoStop}%
\bibitem [{\citenamefont {Brody}(2014)}]{Brody:2014}%
  \BibitemOpen
  \bibfield  {author} {\bibinfo {author} {\bibfnamefont {Dorje~C}\ \bibnamefont
  {Brody}},\ }\bibfield  {title} {{\selectlanguage {English}\enquote {\bibinfo
  {title} {Biorthogonal quantum mechanics},}\ }}\href {\doibase
  10.1088/1751-8113/47/3/035305} {\bibfield  {journal} {\bibinfo  {journal} {J.
  Phys. Math. Theor.}\ }\textbf {\bibinfo {volume} {47}},\ \bibinfo {pages}
  {035305} (\bibinfo {year} {2014})}\BibitemShut {NoStop}%
\bibitem [{\citenamefont {Peierls}(1959)}]{Peierls:1959}%
  \BibitemOpen
  \bibfield  {author} {\bibinfo {author} {\bibfnamefont {R~E}\ \bibnamefont
  {Peierls}},\ }\bibfield  {title} {\enquote {\bibinfo {title} {Complex
  {{Eigenvalues}} in {{Scattering Theory}}},}\ }\href {\doibase
  10.1098/rspa.1959.0176} {\bibfield  {journal} {\bibinfo  {journal} {Proc. R.
  Soc. Math. Phys. Eng. Sci.}\ }\textbf {\bibinfo {volume} {253}},\ \bibinfo
  {pages} {16--36} (\bibinfo {year} {1959})}\BibitemShut {NoStop}%
\bibitem [{\citenamefont {Zworski}(1999)}]{Zworski:1999}%
  \BibitemOpen
  \bibfield  {author} {\bibinfo {author} {\bibfnamefont {Maciej}\ \bibnamefont
  {Zworski}},\ }\bibfield  {title} {\enquote {\bibinfo {title} {Resonances in
  physics and geometry},}\ }\href@noop {} {\bibfield  {journal} {\bibinfo
  {journal} {Not. Am. Math. Soc.}\ }\textbf {\bibinfo {volume} {46}},\ \bibinfo
  {pages} {319--328} (\bibinfo {year} {1999})}\BibitemShut {NoStop}%
\bibitem [{\citenamefont {Mahaux}\ and\ \citenamefont
  {Weidenm{\"u}ller}(1969)}]{Mahaux:1969}%
  \BibitemOpen
  \bibfield  {author} {\bibinfo {author} {\bibfnamefont {Claude}\ \bibnamefont
  {Mahaux}}\ and\ \bibinfo {author} {\bibfnamefont {Hans~A}\ \bibnamefont
  {Weidenm{\"u}ller}},\ }\href@noop {} {\emph {\bibinfo {title} {Shell-Model
  Approach to Nuclear Reactions}}}\ (\bibinfo  {publisher} {{North-Holland Pub.
  Co.}},\ \bibinfo {year} {1969})\BibitemShut {NoStop}%
\bibitem [{\citenamefont {Livsic}(1973)}]{Livsic:1973}%
  \BibitemOpen
  \bibfield  {author} {\bibinfo {author} {\bibfnamefont {Mikhail~S}\
  \bibnamefont {Livsic}},\ }\href@noop {} {\emph {\bibinfo {title} {Operators,
  Oscillations, Waves. ({{Open}} Systems)}}}\ (\bibinfo  {publisher} {{American
  Mathematical Society}},\ \bibinfo {year} {1973})\BibitemShut {NoStop}%
\bibitem [{\citenamefont {Fyodorov}\ and\ \citenamefont
  {Sommers}(2000)}]{Fyodorov:2000}%
  \BibitemOpen
  \bibfield  {author} {\bibinfo {author} {\bibfnamefont {Y~V}\ \bibnamefont
  {Fyodorov}}\ and\ \bibinfo {author} {\bibfnamefont {H~J}\ \bibnamefont
  {Sommers}},\ }\bibfield  {title} {\enquote {\bibinfo {title} {Spectra of
  {{Random Contractions}} and {{Scattering Theory}} for {{Discrete}}-{{Time
  Systems}}},}\ }\href {\doibase 10.1134/1.1335121} {\bibfield  {journal}
  {\bibinfo  {journal} {J. Exp. Theor. Phys. Lett.}\ }\textbf {\bibinfo
  {volume} {72}},\ \bibinfo {pages} {422--426} (\bibinfo {year}
  {2000})}\BibitemShut {NoStop}%
\bibitem [{\citenamefont {Suh}\ \emph {et~al.}(2004)\citenamefont {Suh},
  \citenamefont {Wang},\ and\ \citenamefont {Fan}}]{Suh:2004}%
  \BibitemOpen
  \bibfield  {author} {\bibinfo {author} {\bibfnamefont {Wonjoo}\ \bibnamefont
  {Suh}}, \bibinfo {author} {\bibfnamefont {Zheng}\ \bibnamefont {Wang}}, \
  and\ \bibinfo {author} {\bibfnamefont {Shanhui}\ \bibnamefont {Fan}},\
  }\bibfield  {title} {\enquote {\bibinfo {title} {Temporal coupled-mode theory
  and the presence of non-orthogonal modes in lossless multimode cavities},}\
  }\href {\doibase 10.1109/JQE.2004.834773} {\bibfield  {journal} {\bibinfo
  {journal} {IEEE J. Quantum Electron.}\ }\textbf {\bibinfo {volume} {40}},\
  \bibinfo {pages} {1511--1518} (\bibinfo {year} {2004})}\BibitemShut {NoStop}%
\bibitem [{\citenamefont {Fan}\ \emph {et~al.}(2003)\citenamefont {Fan},
  \citenamefont {Suh},\ and\ \citenamefont {Joannopoulos}}]{Fan:2003}%
  \BibitemOpen
  \bibfield  {author} {\bibinfo {author} {\bibfnamefont {Shanhui}\ \bibnamefont
  {Fan}}, \bibinfo {author} {\bibfnamefont {Wonjoo}\ \bibnamefont {Suh}}, \
  and\ \bibinfo {author} {\bibfnamefont {J~D}\ \bibnamefont {Joannopoulos}},\
  }\bibfield  {title} {\enquote {\bibinfo {title} {Temporal coupled-mode theory
  for the {{Fano}} resonance in optical resonators},}\ }\href {\doibase
  10.1364/JOSAA.20.000569} {\bibfield  {journal} {\bibinfo  {journal} {Opt.
  Soc. Am. J.}\ }\textbf {\bibinfo {volume} {20}},\ \bibinfo {pages} {569--572}
  (\bibinfo {year} {2003})}\BibitemShut {NoStop}%
\bibitem [{Note30()}]{Note30}%
  \BibitemOpen
  \bibinfo {note} {Exponential amplification cannot be indefinite; in practice,
  gain will saturate nonlinearly~\cite {Malzard:2018a}}\BibitemShut {NoStop}%
\bibitem [{\citenamefont {Malzard}\ \emph {et~al.}(2018)\citenamefont
  {Malzard}, \citenamefont {Cancellieri},\ and\ \citenamefont
  {Schomerus}}]{Malzard:2018a}%
  \BibitemOpen
  \bibfield  {author} {\bibinfo {author} {\bibfnamefont {Simon}\ \bibnamefont
  {Malzard}}, \bibinfo {author} {\bibfnamefont {Emiliano}\ \bibnamefont
  {Cancellieri}}, \ and\ \bibinfo {author} {\bibfnamefont {Henning}\
  \bibnamefont {Schomerus}},\ }\bibfield  {title} {\enquote {\bibinfo {title}
  {Topological dynamics and excitations in lasers and condensates with
  saturable gain or loss},}\ }\href {\doibase 10.1364/OE.26.022506} {\bibfield
  {journal} {\bibinfo  {journal} {Opt. Express}\ }\textbf {\bibinfo {volume}
  {26}},\ \bibinfo {pages} {22506} (\bibinfo {year} {2018})},\ \Eprint
  {http://arxiv.org/abs/1705.06895} {arXiv:1705.06895} \BibitemShut {NoStop}%
\end{thebibliography}%

\onecolumngrid
\clearpage
\let\subsection\appendixsection
%
\begin{center}
\textbf{\large Supplemental Material:
\\Bulk-boundary correspondence for non-Hermitian Hamiltonians
\\via Green functions}
\\[0.4ex] Heinrich-Gregor Zirnstein, Gil Refael, and Bernd Rosenow
\end{center}
\par
\setcounter{page}{1}
\twocolumngrid
%
\subsection{Lattice models}
\label{appendix-lattice-models}
In this section, we present the topological phase diagrams for periodic and open boundary conditions for several lattice models with chiral symmetry, and discuss the spatial growth of their bulk Green function.

\begin{figure}
\includegraphics[width=0.46\textwidth]{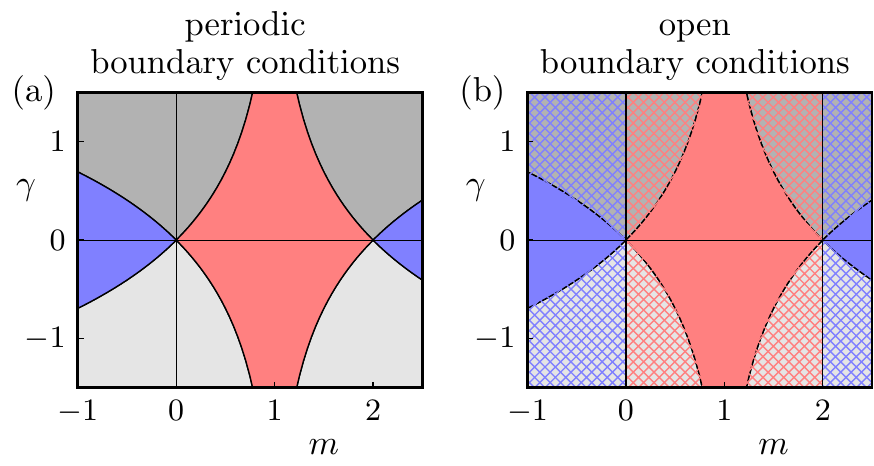}\\
\includegraphics[width=0.46\textwidth]{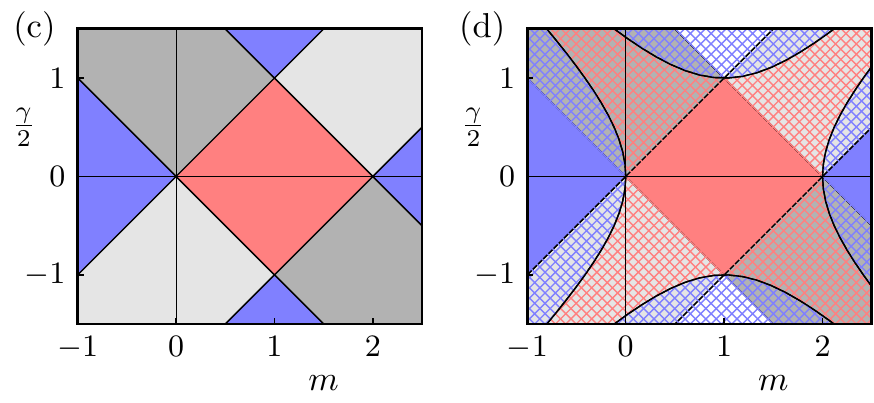}%
\caption{%
Topological phases of non-Hermitian SSH models.
For periodic boundary conditions, (a) and (c), solid colors distinguish four different combinations of the two winding numbers. Different grey tones represent different non-Hermitian phases, where the bulk Green function grows spatially.
For open boundary conditions, (b) and (d), red (blue) indicates the presence (absence) of boundary eigenstates at zero energy.
Around the point $(m,\gamma)=(0,0)$, both diagrams agree with the phase diagram of the continuum model from the main text [Fig.~\ref{fig-phases}].
(a-b) Lattice model Eq.~\eqref{eq-ssh-1}.
(c-d) Lattice model Eq.~\eqref{eq-ssh-2}.
\label{fig-phases-ssh}}
\end{figure}

A non-Hermitian extension of the Su-Schrieffer-Heeger (SSH) model is given by Eq.~\eqref{eq-bloch} with
\begin{equation}
    q_{\pm}(k) = (m-1) + e^{\mp i(k- i \gamma)}
\label{eq-ssh-1}
.\end{equation}
The paths $q_{\pm}(k)$ in the complex plane are circles with center $(m-1)$ and radius $e^{\mp\gamma}$. They wind around the origin whenever the radius exceeds the distance of the center from the origin.
For open boundary condition, the model can be mapped to the Hermitian SSH Hamiltonian $\tilde{H} = S^{-1}HS$ with the same mass parameter by the similarity transform $(S\psi)_j = e^{j\gamma} \psi_j$ where $\psi_j$ denotes the wave function at site $j$. Thus, the presence or absence of boundary eigenstates at zero energy is determined solely by $m$. The topological phase diagram is shown in Fig.~\ref{fig-phases-ssh}(a).

In the literature, a different non-Hermitian extension of the SSH model has been discussed~\cite{Lee:2016,Kunst:2018,Yao:2018b,Xiong:2018,Lee:2019c,Herviou:2019}.
It is given by Eq.~\eqref{eq-bloch} with
\begin{equation}
    q_{\pm}(k) = (m-1) \pm \gamma/2 + e^{\mp ik}
\label{eq-ssh-2}
.\end{equation}
The paths are circles with unit radius, but their center is shifted by a distance $\gamma/2$ along the real axis.
For open boundary conditions, the model can be mapped to a Hamiltonian $\tilde{H} = S^{-1}HS$ with off-diagonal entries
\begin{equation}
    \tilde{q}_{\pm}(k) = \sqrt{(m-1)^2 - (\gamma/2)^2} + e^{\mp ik}
.\end{equation}
This Hamiltonian is Hermitian if $|m-1|>|\gamma/2|$; this region includes the point $(m,\gamma)=(0,0)$ that features prominently in the continuum model.
Here, the similarity transform is given by $(S\psi)_j = r^j \diag(1,r) \psi_j$ where $\diag$ denotes a $2\times 2$-diagonal matrix and $r = \sqrt{[(m-1)-\gamma/2]/[(m-1)+\gamma/2]}$~\cite{Yao:2018b}.
Thus, the boundary eigenstates eigenstates at zero energy are determined by a Hermitian SSH model with different parameters. For the region where this transformation does not yield a Hermitian Hamiltonian, we refer to Ref.~\cite{Yao:2018b}. The phases are shown in Fig.~\ref{fig-phases-ssh}(b).

We have indicated that if the non-Hermitian winding number is nonzero, then the bulk Green function grows exponentially.
For lattice models, this can be justified as follows: If the model has short-range hopping, then the Hamiltonian is a (Laurent) polynomial $H(k)=\tilde{H}(z)$ in the variables $z=e^{ik}$ and $z^{-1}=e^{-ik}$. In the generic case, the bulk Green function can be written as a sum of exponentials $z_s^j$, where $j$ denotes a lattice site, and the $z_s$ are the complex zeros of $\det \tilde{H}(z)$.
Each zero inside the unit circle contributes $+1$, while each zero outside the unit circle contributes $-1$ to the non-Hermitian winding number $\nu(H) = (2\pi)^{-1} \int_{0}^{2\pi} dk\, \partial_k \arg\det(H(k))$. (There are also contributions from a pole at $z=0$.)
Thus, the winding number changes when one of the zeros crosses the unit circle, i.e.\ when $|z_s|<1$ changes to $|z_s|>1$, or vice versa; this means a change in spatial growth.
We refer to the companion paper Ref.~\cite{Zirnstein:2020} for a full proof.
That said, if two zeros cross the unit circle simultaneously, then it is possible that no change in spatial growth occurs; this happens, for example, when the sign of the mass is changed in the Hermitian SSH model.

The converse statement is not true: It may happen that the non-Hermitian winding number is zero, but the bulk Green function grows exponentially. This occurs for the model \eqref{eq-ssh-2} in the parameter region $|\gamma/2|>|m-1|+1$ [Fig~\ref{fig-phases-ssh}(c-d)].
If we allow the parameter $m$ to become complex, thus leaving the BDI symmetry class while remaining in AIII, spatial growth may happen even when the Bloch Hamiltonian can be deformed to a Hermitian Bloch Hamiltonian without crossing zero energy. Setting $m=1 + te^{i\phi}$ where $t>0$ and $\phi$ is real, we find, for $\gamma=0$,
\begin{equation}
    q_{\pm}(k) = t e^{i\phi} + e^{\mp ik}
.\end{equation}
Changing $\phi$ continuously deforms the corresponding Hamiltonian; for $\phi=0$, the Hamiltonian is Hermitian and the bulk Green function decays in space, while for $\phi=\pi/2$, the Hamiltonian is non-Hermitian and features gain $+it$.
It is plausible that if $t \gg 1$, then the bulk Green function grows exponentially.
In more detail, the zeros of $\det(\tilde{H}(z)-i\Gamma)$ are
\begin{equation}
    z_{1,2}(i\Gamma) = \frac{i}{2}\left[\frac{1+\Gamma^2 - t^2}{t} \pm \sqrt{\frac{(1+\Gamma^2-t^2)^2}{t^2}+4}\right]
,\end{equation}
where $\Gamma>0$ denotes an additional dissipation.
During analytic continuation from $\Gamma=+\infty$ to $\Gamma=0$, both zeros cross the unit circle if and only if $t>1$, which means that the bulk Green function grows exponentially in this parameter regime~\cite{Zirnstein:2020}.

\subsection{Phenomenological justification of the Mahaux-Weidenm\"{u}ller formula}
\label{appendix-mahaux-pheno}
In this section, we present a phenomenological justification of the Mahaux-Weidenm\"{u}ller formula, Eq.~\eqref{eq-scattering-matrix}, for the scattering setup illustrated in Fig.~\ref{fig-scattering}(a) of the main text.

The Mahaux-Weidenm\"{u}ller formula calculates the scattering matrix of a system, which describes how a monochromatic incoming amplitude $\phi^-$ at energy $E$ is mapped to an outgoing amplitude $\phi^+ = S(E)\phi^-$. We repeat it here for convenience:
\begin{equation}
    S(E)
    = \Id - 2i W^\dagger \frac1{E - \hat H_0 + iWW^\dagger} W
\label{eq-appendix-mahaux}
.\end{equation}
The system is described by the Hamiltonian $\hat H_0$, while the matrix $W$ maps the outside field to a state inside the system, and vice versa for $W^\dagger$.
Here, the matrix products are to be understood as integrals over space and sums over internal degrees of freedom. For example, $W \phi^- \equiv \sum_{\sigma}\int dx_1\, W_{\tau\sigma}(x,x_1)\phi_{\sigma}^-(x_1)$, where $x_1,x$ denote positions, and $\tau,\sigma$ internal degrees of freedom.

The formula can be justified by the following ansatz:~\cite{Fyodorov:2000,Fan:2003,Suh:2004}
Let us denote the state of the internal system by $\psi$. If the system were isolated, then this state would evolve according to the linear Schr\"{o}dinger equation $i\partial_t \psi = \hat H_0 \psi$. If the Hamiltonian $\hat H_0$ is Hermitian, then the square integral of the state is conserved, $\partial_t(\psi^\dagger \psi)=0$; this corresponds to the conservation of probability in quantum mechanics, or energy in optics.
To couple the system to the outside field, we now make a linear ansatz%
\begin{subequations}
\begin{align}
    i\partial_t \psi &= A \psi + B \phi^-
    \\
    \phi^+ &= C \psi + D \phi^-
,\end{align}
\end{subequations}
with some matrices $A,B,C,D$, which we assume to be constant in time.
The conservation law now becomes
\begin{equation}
    \partial_t(\psi ^\dagger \psi ) = (\phi ^-)^\dagger \phi ^- - (\phi ^+)^\dagger (\phi ^+)
,\end{equation}
which means that the incoming and outgoing amplitudes may add to or remove from the conserved quantity of the system state. In optics, this ansatz is known as temporal coupled-mode theory~\cite{Suh:2004,Fan:2003}.
After some algebra, we obtain that the combination of the ansatz and the conservation law is equivalent to the equations
\begin{subequations}
\begin{align}
    i\partial _t \psi &= \hat H_0 \psi - \frac1{\sqrt{2}}W(\phi ^+ + S_0\phi ^-) \\
    (\phi ^+ - S_0\phi ^-) &= i\sqrt{2}W^\dagger \psi 
,\end{align}
\end{subequations}
and the constraints that $\hat H_0$ be Hermitian and $S_0$ be unitary.
The first equation can be viewed as a Schr\"{o}dinger equation with a source term, and the second as a constraint that relates the outside field and the system.
We proceed by making a change of basis for $S_0\phi^- \to \phi^-$, so that we can assume that $S_0$ is the identity matrix. Then, we make the ansatz of a harmonic motion $\psi(t) = e^{-iEt}\psi(E)$ and eliminate $\psi$ from the equations and the Mahaux-Weidenm\"{u}ller formula follows.

The ansatz is phenomenological.
In real systems, the coupling $W$ may be energy-dependent, and instead of focusing the outside field to a single point $x_1$, one may have to consider plane waves incident at different angles. We present a more microscopic derivation in the next section.

\subsection{Microscopic derivation of the\\ Mahaux-Weidenm\"{u}ller formula}
\label{appendix-mahaux-micro}
In this section, we present a self-contained derivation of the Mahaux-Weidenm\"{u}ller formula, Eq.~\eqref{eq-scattering-matrix}, for a microscopic model of the scattering setup illustrated in Fig.~\ref{fig-scattering}(a) of the main text. In the setup, we label the horizontal direction by $x$ and the vertical direction by $z$.

For simplicity, we will model the outside field as a scalar field $\phi (t,x,z)$ that propagates according to the wave equation, $(\partial _t^2-\Delta )\phi =0$.
A monochromatic configuration with energy $E$ that satisfies the wave equation can be expanded in terms of ingoing amplitudes $\phi _\alpha ^+$ and outgoing amplitudes $\phi _\alpha ^-$:
\begin{align}
    \phi _E(t,x,z)
    &= e^{-iEt} \left[
    \sum_\alpha  \left[ \phi _\alpha ^+ e^{ik_z^\alpha z} + \phi _\alpha ^- e^{-ik_z^\alpha z}\right] u_\alpha (x)
    \right.
    \nonumber
    \\ &\quad
    \left.
    + \sum_\beta \phi _\beta ^+ e^{-\kappa ^\beta z}u_\beta (x)
    \right]
    \label{eq-phi-expansion}
.\end{align}
Here, the functions $u_\alpha (x)$ are eigenfunctions of the one-dimensional Laplacian, $u_\alpha (x)\propto \exp(ik_x^\alpha x)$ where $\alpha $ labels the possible horizontal momenta $k_x^\alpha $. When this momentum is not too large, the field can propagate in $z$-direction with wave vector $k_z^\alpha = \sqrt{E^2 - (k_x^\alpha )^2}$. This is known as an open scattering channel. On the other hand, if the horizontal momentum is larger than the available energy, the scattering channel is closed, and only an evanescent wave with decay $\kappa _z^\beta = \sqrt{(k_x^\alpha )^2 - E^2}$ remains. For notational clarity, we reserve the label $\beta $ for this case.

The scattering problem now asks to calculate the outgoing amplitudes $\phi _\alpha ^+$ from the incoming amplitudes $\phi _\alpha ^-$. The solution is the scattering matrix $S(E)$, which gives $\phi ^\alpha _+ = \sum_{\alpha \tilde{\alpha }} S_{\alpha \tilde{\alpha }}(E) \phi _{\tilde{\alpha }}^-$.

To do that, we need to specify the dynamics of the internal system and the coupling to the outside.
We have already indicated that internal system is located at $z=0$, and its state represented by a function $\psi (t,x)$ that evolves according to a Schr\"{o}dinger equation $i\partial _t\psi = \hat H_0\psi $ with a Hermitian Hamiltonian $\hat H_0$.
The most economic way to specify the whole setup is to write a Lagrangian $L=L_\phi + L_\psi + L_V$ with terms
\begin{subequations}
\begin{align}
    L_\phi &= \frac12 \int dx\,\int_{0}^\infty dz\, [(\partial _t \phi ^*)(\partial _t \phi ) - (\nabla \phi ^*)\cdot (\nabla \phi )] \\
    L_\psi &= \int dx\, \psi ^\dagger (t,x)(i\partial _t - \hat H_0)\psi (t,x) \\
    L_V &= \psi ^\dagger V \phi |_{z=0} + \phi ^*|_{z=0}V^\dagger \psi 
.\end{align}
\end{subequations}
The first terms gives the wave equation, the second term the Schr\"{o}dinger equation, and the third term is a potential energy that couples the outside field at the boundary $z=0$ to the internal state via an operator $V$.
We proceed by deriving the equations of motion while paying special attention to the boundary terms at $z=0$, and find
\begin{subequations}
\begin{align}
    \label{eq-motion}
    i\partial _t \psi &= \hat H_0 \psi - V \phi |_{z=0} \\
    \label{eq-wpsi}
    \partial _z \phi |_{z=0} &= -2V^\dagger \psi 
.\end{align}
\end{subequations}
In other words, the values $\phi |_{z=0}$ of the outside field at the boundary act as a source term for the Schr\"{o}dinger equation, while the internal state poses a constraint on the spatial derivative $\partial _z \phi |_{z=0}$ of the outside field.
For a harmonic oscillation, $i\partial _t \to E$, we can solve the equations of motion and obtain the scattering matrix
\begin{equation}
    \label{eq-scattering}
    S_{\alpha \tilde{\alpha }}(E) =
        \delta_{\alpha \tilde{\alpha }} - \sum_{i,j} \frac{4i}{k_z^\alpha } (V^+)_{\alpha i}
        \left[\frac{1}{E - H_{\text{eff}}}\right]_{ij} V_{j\tilde{\alpha }}
,\end{equation}
where $i,j$ label orthonormal states of the state $\psi $, and
\begin{equation}
    H_{\eff,ij}
    = \hat H_{0,ij}
    -\sum_{\beta } \frac{2}{\kappa ^\beta } V_{i\beta }(V^+)_{\beta j}
    -i \sum_{\alpha } \frac{2}{k_z^\alpha }  V_{i\alpha }(V^+)_{\alpha j}
\end{equation}
is an effective Hamiltonian.
It contains both additional Hermitian terms that arise from the closed scattering channels, and additional antihermitian terms that arise from the open scattering channels.
To obtain the Mahaux-Weidenm\"{u}ller formula in the phenomenological form presented in the main text, we have to perform several simplifications: (a) the Hermitian Hamiltonian $\hat H_0$ is redefined to incorporate the potential energy due to the closed channels, (b) the matrix $W_{i\alpha }$ is obtained from $V_{i\alpha }$ by normalizing the scattering amplitudes with $\sqrt{k_z^\alpha }$, (c) the energy dependence of $k_z^\alpha $ is neglected, making $W_{i\alpha }$ independent of energy, and (d) transforming the Fourier basis $u_\alpha (x)$ back to the position basis, if desired.

\end{document}